\documentclass[twocolumn]{aastex631}

\usepackage{amsmath}

\shorttitle{Soft X-ray Flares on Solar-type Stars}
\shortauthors{Zhao et al.}

\graphicspath{{./}{figures/}}

\begin{document}

\title{A Statistical Study of Soft X-ray Flares on Solar-type Stars}

\author{Z. H. Zhao}
\affiliation{School of Astronomy and Space Science, Nanjing University, Nanjing 210023, People's Republic of China; xincheng@nju.edu.cn}
\affiliation{Key Laboratory of Modern Astronomy and Astrophysics (Nanjing University), Ministry of Education, Nanjing 210093, People's Republic of China}

\author{Z. Q. Hua}
\affiliation{School of Astronomy and Space Science, Nanjing University, Nanjing 210023, People's Republic of China; xincheng@nju.edu.cn}
\affiliation{Key Laboratory of Modern Astronomy and Astrophysics (Nanjing University), Ministry of Education, Nanjing 210093, People's Republic of China}

\author{X. Cheng}
\affiliation{School of Astronomy and Space Science, Nanjing University, Nanjing 210023, People's Republic of China; xincheng@nju.edu.cn}
\affiliation{Key Laboratory of Modern Astronomy and Astrophysics (Nanjing University), Ministry of Education, Nanjing 210093, People's Republic of China}

\author{Z. Y. Li}
\affiliation{School of Astronomy and Space Science, Nanjing University, Nanjing 210023, People's Republic of China; xincheng@nju.edu.cn}
\affiliation{Key Laboratory of Modern Astronomy and Astrophysics (Nanjing University), Ministry of Education, Nanjing 210093, People's Republic of China}

\author{M. D. Ding}
\affiliation{School of Astronomy and Space Science, Nanjing University, Nanjing 210023, People's Republic of China; xincheng@nju.edu.cn}
\affiliation{Key Laboratory of Modern Astronomy and Astrophysics (Nanjing University), Ministry of Education, Nanjing 210093, People's Republic of China}

\correspondingauthor{X. Cheng}
\email{xincheng@nju.edu.cn}

\begin{abstract}

The statistical characteristic of stellar flares at optical bands has received an extensive study, but it remains to be studied at soft X-ray bands, in particular for solar-type stars. 
Here, we present a statistical study of soft X-ray flares on solar-type stars, which can help understand multi-wavelength behaviors of stellar flares. 
We mainly use Chandra Source Catalog Release 2.0, which includes a number of flaring stars with denoted variability, and Gaia Data Release 3, which includes necessary information for classifying stars. We also develop a set of methods for identifying and classifying stellar soft X-ray flares and estimating their properties. 
A detailed statistical investigation for 129 flare samples on 103 nearby solar-type stars as selected yields the following main results. 
(1) The flare energy emitted at the soft X-ray band in our sample ranges from $\sim 10^{33}$ to $\sim 10^{37} \ \mathrm{erg}$, and the majority of them are superflares with the most energetic one having energy of $6.0_{-4.7}^{+3.2} \times 10^{37} \ \mathrm{erg}$. 
(2) The flare duration is related to its energy as formulated by $T_\mathrm{duration,SXR} \propto E_\mathrm{flare,SXR}^{\ 0.201 \pm 0.024}$, which is different from those derived at optical and NIR bands, indicating distinct radiation mechanisms at different bands. 
(3) The frequency distribution of stellar flares as a function of energy is formulated as $\mathrm{d} N_\mathrm{flare} / \mathrm{d} E_\mathrm{flare,SXR} \propto E_\mathrm{flare,SXR}^{\ -1.77}$, which is similar to the results found at other bands and on other types of stars, indicating that the energy emitted at the soft X-ray band could be a constant fraction of the full-band bolometric energy. 

\end{abstract}



\section{Introduction}
\noindent Stellar flares, which are intense energy-emitting phenomena taking place in the atmosphere of stars, are 
commonly argued to be caused by magnetic reconnection \citep[e.g., ][]{Gershberg2005book, Benz2010ARAA48.241, Shibata2016IAUS320.3}. 
Except for long-term activities, which refer to the cyclic evolution of magnetic activities of stars and have likely an impact on the atmospheric temperature and chemical composition changes of planets \citep{Chen2021NA5.298}, stellar flares, along with possible coronal mass ejections (CMEs), may cause disastrous space weather and affect the geomagnetic fields and thus the habitability of the planets orbiting these stars \citep[e.g., ][]{Tsurutani2003JGRA108.1268}. 

\indent With new generations of instruments working at distinct bands, it becomes possible to conduct studies on stellar magnetic activities in detail, such as large-sample time-domain statistical surveys. 
\citet{Maehara2012Nat485.478,Maehara2015EPS67.59} and \citet{Shibayama2013ApJS209.5} statistically studied stellar optical flares from solar-type stars with Kepler data. 
They found that flares on solar-type stars emit much more energy than solar flares, and the power-law index $\gamma$ in the frequency distribution $\mathrm{d}N_\mathrm{flare}/\mathrm{d}E_\mathrm{flare} \propto E_\mathrm{flare}^{\gamma}$ is roughly $-1.8$. With Transiting Exoplanet Survey Satellite (TESS) data, \cite{Doyle2020MNRAS494.3596} and \citet{Tu2020ApJ890.46,Tu2021ApJS253.35} further confirmed the conclusions drawn with Kepler data. 
\cite{Pye2015AA581.A28} studied soft X-ray (SXR) flares from Tycho cool (F-M type) stars with XMM-Newton data and 
\cite{Getman2021ApJ916.32} studied SXR flares from pre-main sequence (PMS) stars with Chandra data. 
It is interesting that stellar SXR flares show a similar power-law index $\gamma$, i.e. about $-1.8$, to optical flares in the frequency distribution. 
Such a law is also true for different types of stars, indicating that stellar flares may all share the same energy release process, i.e., magnetic reconnection. 

\indent Based on the magnetic reconnection theory, \citet{Maehara2015EPS67.59} derived that the bolometric duration of stellar flares is well associated with their energy, which can be formulated as $T_\mathrm{duration,bol} \propto E_\mathrm{flare,bol}^{\sim 1/3}$, 
in good agreement with solar white-light flares \citep[e.g., ][]{Namekata2017ApJ851.91}. 
Such a correlation was subsequently verified by stellar optical and near-infrared (NIR) observations \citep[e.g., ][]{Maehara2015EPS67.59,Tu2020ApJ890.46,Tu2021ApJS253.35}. 
However, it is unclear if such a power law is still valid and what is the slope if using flares observed on solar-type stars at the SXR band. 
Moreover, stellar flares often show different properties at different bands including the distributions of their duration and energy, thus it is also necessary to compare the derived statistical properties at the SXR band with that at the white-light band. 




\indent Flares could be accompanied with coronal mass ejections (CMEs) \citep[e.g.][]{Yashiro2009IAUS257.233,Aarnio2011SoPh268.195}, which are defined as large-scale expulsions of magnetized plasma from stellar atmosphere \citep[e.g.][]{Chen2011LRSP8.1,Veronig2021NA5.697,Lu2022AA663.A140}. 
Studies on solar flares revealed that such an association holds better for stronger and/or longer-duration \citep[e.g.][]{Cheng2010ApJ712.752} flares, 
which indicates that the properties of stellar flares could be used to search for and diagnose potential CMEs. 
Type II radio bursts, i.e., slowly-drifting radio emission features, are also valuable for identifying and constraining the properties of stellar CMEs. 
By employing this approach, \citet{Crosley2018ApJ856.39} performed a multi-wavelength analysis of an M dwarf EQ Peg, and predicted the CME parameters according to solar scaling relationships. 
Very interestingly, \citet{Veronig2021NA5.697} recently provided a strong indication for stellar CMEs based on the detection of coronal dimmings on stars. 
\citet{Loyd2022ApJ936.170} also performed a coronal dimming analysis on a young K2 dwarf $\epsilon$ Eri and proposed that coronal dimmings can be used to constrain CME masses.

\indent 
In order to thoroughly understand the nature of flares on solar-type stars, in this work, we conduct a detailed statistical study based on a combination of Chandra Source Catalog Release 2.0 (CSC 2.0) and Gaia Data Release 3 (Gaia DR3). 
The derived statistical properties are also expect to be able to help detect possible associated-coronal mass ejections. 
In Section \ref{section_data_and_methods}, we first develop a set of methods for identifying and classifying stellar SXR flares and calculating their parameters. 
The results including flare properties and their correlations, as well as occurence frequency distributions, are shown in Section \ref{section_results_and_discussions}, in which we also make a comparison between stellar and solar flares. 
A brief summary is given in Section \ref{section_summary}.

\section{Data and methods} \label{section_data_and_methods}


\subsection{Selection of flare candidates} \label{section_selection_of_flare_candidates}
\begin{figure*}
	\centering
	\includegraphics[width=0.6\textwidth]{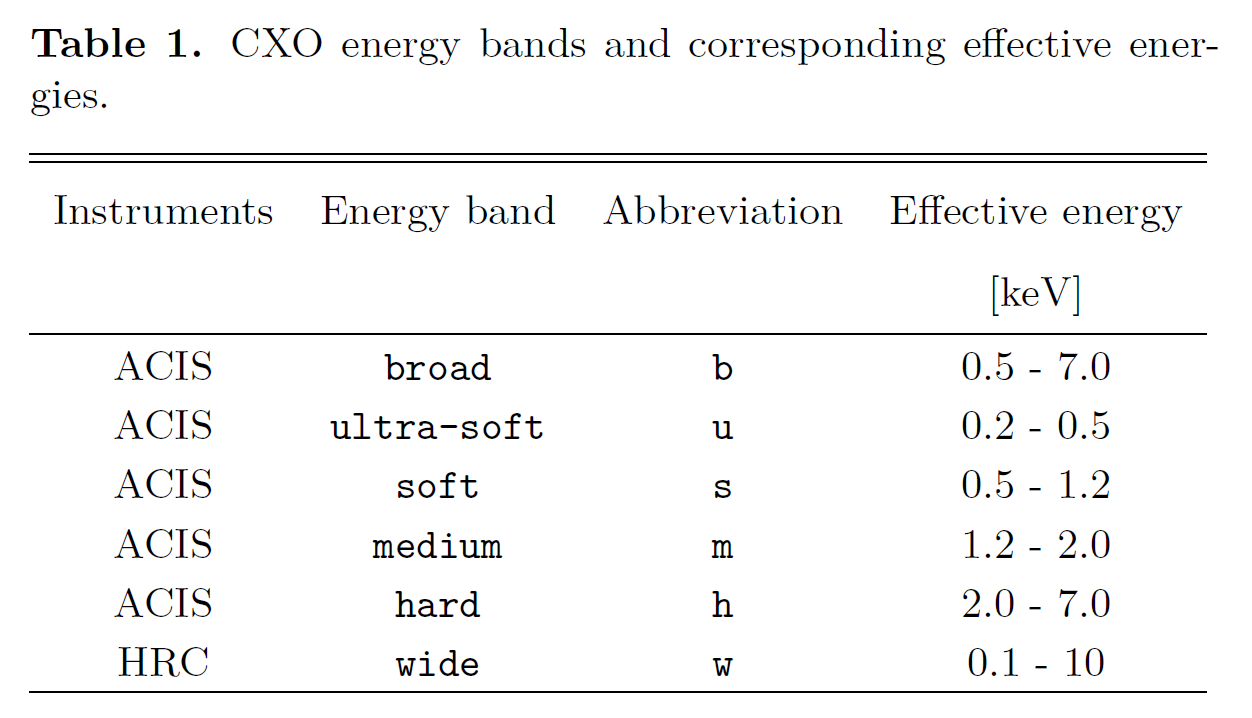}
	\caption{Table 1}
	\label{CXO_bands}
\end{figure*}

\noindent We search for stellar flares using data in the Chandra Source Catalog Release 2.0 (CSC 2.0)\footnote{\url{https://cxc.cfa.harvard.edu/csc}} collected by NASA's Chandra X-ray Observatory (CXO)\footnote{\url{https://cxc.harvard.edu}} through the end of 2014. 
CXO has an unprecedented spatial resolution of subarcsecond for most detected sources, which allows us to match their optical counterparts unambiguously. 
Our primary database, CSC 2.0, is the second major release of the definitive catalog of X-ray sources detected by the CXO. 
CSC 2.0 provides properties for $317,167$ individual X-ray sources and $928,280$ individual observation detections, allowing statistical analysis of large samples and also individual case studies. 
There are information about each source across 5 science energy bands (\verb|broad|, \verb|hard|, \verb|medium|, \verb|soft|, and \verb|ultra-soft|) for Advanced CCD Imaging Spectrometer (ACIS) and 1 band (\verb|wide|) for High Resolution Camera (HRC) \footnote{\url{https://cxc.cfa.harvard.edu/csc/columns/ebands.html}}. 
All of the 6 science energy bands and their effective energies are given in Figure \ref{CXO_bands}. 

\indent In this work, we search for stellar flare candidates from the ACIS data by using a parameter called Variability Index (\verb|var_index|) available in CSC 2.0. 
Variability Index is an integer in the range of $[0,10]$ which is calculated for each science energy band. The calculation method combines the Gregory-Loredo variability probability with the fractions of the multi-resolution light curve output by the Gregory-Loredo analysis that are within $3\sigma$ and $5\sigma$ of the average count rate 
\footnote{\url{https://cxc.cfa.harvard.edu/csc/why/gregory\_loredo.html}}. 
According to the statements on the websites of CSC 2.0, we evaluate whether the source region flux is uniform throughout the observation out of the value of this parameter. The events in CSC 2.0 with $ \verb|var_index| \geq 6 $ are commented as a definite variable, which is used for identifying possible flaring stars. 

\indent We first select the sources observed by ACIS with $\verb|var_index| \geq 6$ 
in at least one ACIS or HRC band, and obtained 35,212 events. 
We cross-match these events with Gaia Data Release 3 
(Gaia DR3)\footnote{\url{https://www.cosmos.esa.int/web/gaia/data-release-3}} catalog, using a relatively small radius of $1''$ in order to restrain the proportion of false-matched results. 
The cross-matching process is done by using the TOPCAT\footnote{\url{http://www.star.bris.ac.uk/\%7Embt/topcat}}. 
When doing the cross-matching, the proper motion of the stars can be neglected because none of the sources have a total proper motion greater than the error of the angular distance between the two catalogs, which is calculated from the coordinates given by the two catalogs and their errors. 
Among the Gaia DR3 counterparts, there are 8,688 events with credible distances, i.e. the parallaxes are available and the relative parallax errors are lower than $0.2$, given that the distance is a critical parameter when calculating the fluence or energy of a given stellar flare event. 
With parallaxes $p$ and parallax errors $\sigma_p$, one can calculate the distances $d$ and distance errors $\sigma_d$ of the sources. 

\indent Next, we try to exclude astrometrically or photometrically suspectable sources, by using Gaia DR3 parameters \verb|astrometric_excess_noise|, 
which characterises the goodness of fit of the astrometric solution to the observations, 
and \verb|phot_bp_rp_excess_factor|, 
which is related to photometric reliability. 
We require \verb|astrometric_excess_noise| to be lower than 1, and 
\verb|phot_bp_rp_excess_factor|, which is color-dependent, to be low enough so that the over-densed region in the  \verb|phot_bp_rp_excess_factor| - \verb|bp_rp| diagram is included and unusually high values are excluded. 
Such a sifting process follows the methods in the Section 4 of \textit{Tutorial: Exploring Gaia data with TOPCAT and STILTS}\footnote{\url{https://github.com/mbtaylor/tctuto}} \citep{Taylor2016tutorial}. 
4,622 CSC events remain after this sifting process. 
These flare candidate events are related to 1,680 independent CXO exposures, 
which are identified with an independent observation identifier (ObsID) each. 
Besides, we download a Gaia DR3 catalog which contains all of the sources within $100 \ \mathrm{pc}$, 
and process it by using the same method mentioned above to exclude astrometrically or photometrically spurious sources. 
With these numerous Gaia sources as a background, we can easily distinguish the positions of our CSC sources on a Hertzsprung–Russell diagram (HRD) and further specify their types.

\subsection{Selection of solar-type stars} \label{section_selection_of_solar-type_stars}
\noindent We used the effective temperature ($T_\mathrm{eff}$) and the surface gravity in logarithmic scale ($\log{g}$) available from Gaia DR3 to select solar-type stars. 
Following the criteria used in previous studies 
\citep[e.g., ][]{Maehara2012Nat485.478,Shibayama2013ApJS209.5,Maehara2015EPS67.59,Tu2020ApJ890.46,Tu2021ApJS253.35}, 
we set the selection criteria as follows: 
(1) $5100 \ \mathrm{K} \leq T_\mathrm{eff} < 6000 \ \mathrm{K}$ and 
(2) $4.0 \leq \log{g} < 4.8 $. 

\indent Out of the 4,622 events, the number of events satisfying the selection criteria, i.e. flare candidates on solar-type stars, is 528. 

\subsection{Analysis of light curves} \label{section_analysis_of_light_curves}
\noindent In this work, we define target regions by encircled counts fraction (ECF). 
We set $r_{90}$ as the radius of each target region, which is defined by a circle around the target source inside which ECF reaches 90\% and above. 
Besides, the background region is defined as an annulus 
with inner radius $2 r_{90}$ and outer radius $3 r_{90}$. 
Photon count rates (denoted as CR in the following parts) in the two regions mentioned above are denoted as $\mathrm{CR}^{\mathrm{tar}}$ and $\mathrm{CR}^{\mathrm{bkg}}$, respectively. 

\indent We then define the real $\mathrm{CR}(t)$ coming from the target source 
in each time bin 
as 
\begin{equation}
    \mathrm{CR}(t) = \frac{1}{0.9} \left[\mathrm{CR}^{\mathrm{tar}}(t) - \frac{1}{5} \mathrm{CR}^{\mathrm{bkg}}(t) \right], 
\end{equation}
where $\mathrm{CR}^{\mathrm{bkg}}(t)$ is multiplied by a factor of 1/5 because the area of the background region is 5 times the target region. 

\begin{figure*} 
	\centering
    \includegraphics[width=0.9\textwidth]{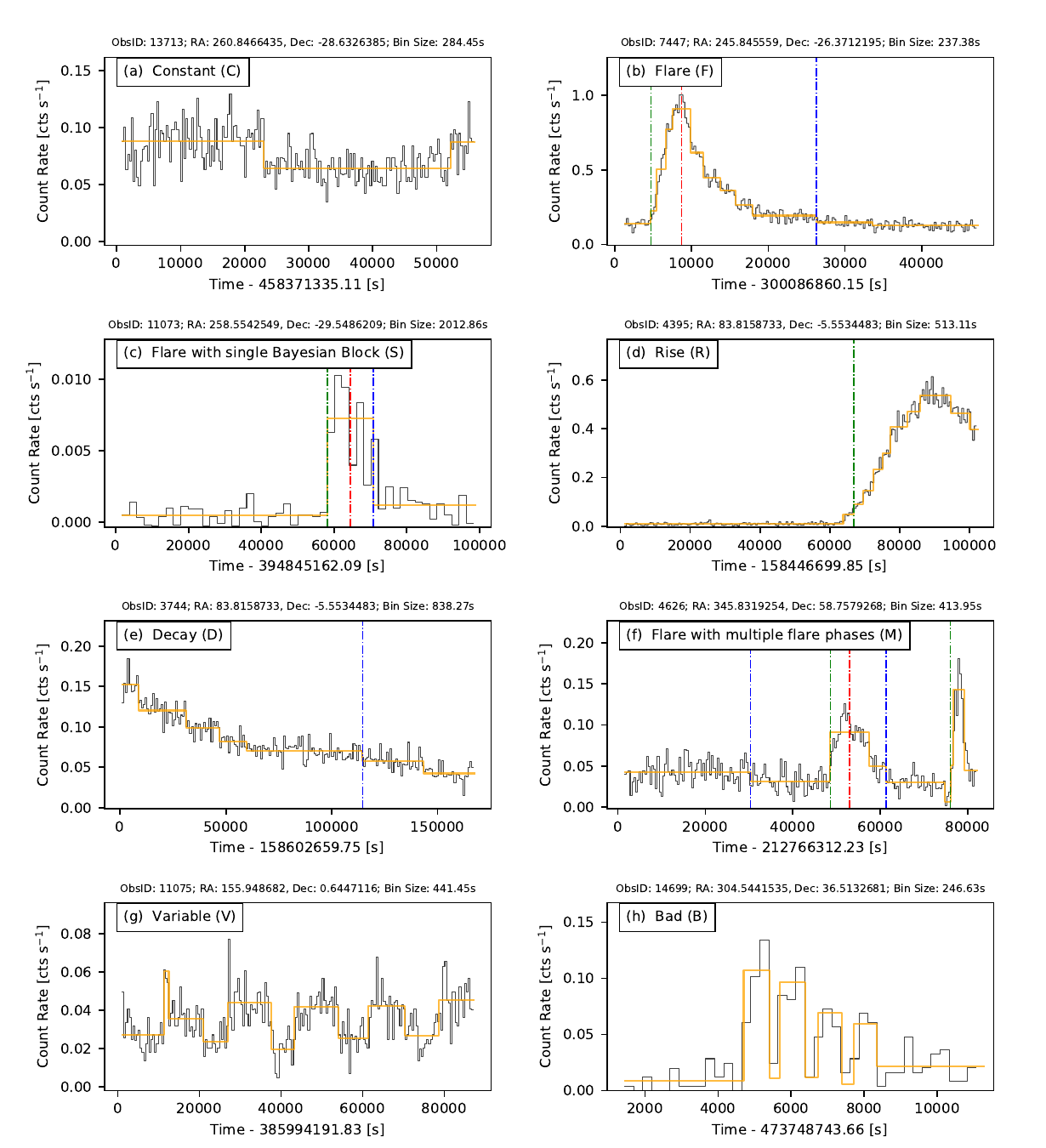}
	\caption{
	Examples of the light curves for the eight categories: 
	(a) Constant (C), (b) Flare (F), (c) Flare with single Bayesian Block (S), (d) Rise (R), (e) Decay (D), 
	(f) Flare with multiple flaring phases (M), (g) Variable (V) and (h) Bad (B). 
 Binned light curves and Bayesian Blocks are shown with black and orange lines, respectively. 
 The observation identifiers (ObsIDs), equatorial coordinates (including RA and Dec) and bin sizes of corresponding light curves are given above each panel. 
 Three time parameters, $t_\mathrm{start}$, $t_\mathrm{peak}$ and $t_\mathrm{end}$, of each flaring phase are illustrated with green, red and blue dash-dotted lines, respectively. 
	}
	\label{lc_examples}
\end{figure*}

\indent We search for stellar flares by inspecting light curves of events from solar-type stars (see Section \ref{section_selection_of_solar-type_stars}) among the 528 candidate events mentioned above. 
In this work, we plot the light curves, the CRs against the arrival times, 
after binning the arrival times with an appropriate time interval, 
which specifically means that 
there should be at most 200 bins in one light curve with no more than $25 \%$ of the bins containing no photons. 
In order to determine the times at which the photon count rate changes obviously and grasp the general tendencies of the light curves, we adopt an approach named Bayesian Blocks, 
which is widely used to identify and characterize flaring behavior in X-ray sources 
\citep[e.g., ][]{Scargle1998ApJ.504.405,Scargle2013ApJ764.167,Getman2021ApJ916.32}. 
After fitting the original light curve with Bayesian Blocks, we get a series of times when the $\mathrm{CR}$ has noticeable changes, i.e. changepoints, as well as a series of stepwise constant $\mathrm{CR}$, as shown in Fig. \ref{lc_examples}. 

\indent Before identifying stellar flares in our data, it is needed to define the quiescent phase and flaring phase in a light curve. 
Here, we set the lowest value of $\mathrm{CR}$ of Bayesian Blocks during an event as the quiescent level (denoted as $\mathrm{CR}_\mathrm{q}$). 
We define a relative count rate $\Delta \mathrm{CR}$ as 
\begin{equation}
    \Delta \mathrm{CR}(t) = \mathrm{CR}(t) - \mathrm{CR}_\mathrm{q}, 
\end{equation}
and then define the start time $t_\mathrm{start}$ and end time $t_\mathrm{end}$ of a flare as the first and last changepoints between which the $\Delta \mathrm{CR}$ of Bayesian Blocks are consecutively higher than 5 times the photometric errors, i.e. 
\begin{equation}
    \Delta \mathrm{CR}(t) > 5 \sigma_\mathrm{CR}, \quad (t_\mathrm{start} < t < t_\mathrm{end}), 
\end{equation}
where $\sigma_\mathrm{CR}$ denotes the photometric error with the same unit as $\mathrm{CR}$. 
The numerical value of 
$\sigma_\mathrm{CR}$ is estimated by the root-mean-square error of $\mathrm{CR}_\mathrm{q}$, i.e. 
\begin{equation}
    \sigma_\mathrm{CR} = \frac{\sqrt{\mathrm{CR}_\mathrm{q} \Delta t_\mathrm{q}}}{\Delta t_\mathrm{q}} = \sqrt{\frac{\mathrm{CR}_\mathrm{q}}{\Delta t_\mathrm{q}}}, 
\end{equation}
where $\Delta t_\mathrm{q}$ is the corresponding time range of $\mathrm{CR}_\mathrm{q}$. 
Then, the quiescent phase is defined as the time interval 
$(t_0,\ t_\mathrm{start}) \cup (t_\mathrm{end},\ t_1)$ for events in categories F or S, or 
$(t_{\mathrm{end}, i},\ t_{\mathrm{start} , i+1})$ as the quiescent phase 
between the $i \mathrm{th}$ and $(i+1) \mathrm{th}$ flaring phases for events in categories M-F or M-S; 
while the flaring phase is defined as the time interval $(t_\mathrm{start},\ t_\mathrm{end})$ for events in categories F or S, or 
$(t_{\mathrm{start}, i},\ t_{\mathrm{end}, i})$ as the $i\mathrm{th}$ flaring phase for events in categories M-F or M-S, 
where $t_0$ and $t_1$ refer to the start and end time of the corresponding CXO exposure respectively. 

\indent All of the light curves and Bayesian Blocks fitted to them are examined both numerically and then visually, 
and classified into eight categories as illustrated in Fig. \ref{lc_examples}: 

\noindent \textbf{(1) Constant (C): }
This category contains events whose light curves have no Bayesian Blocks with $\Delta \mathrm{CR}$ higher than $5 \sigma_\mathrm{CR}$. This commonly occurs when an event is variable but the variability of this event is not high enough, making it difficult to determine whether this event contains a typical stellar flare. 
We exclude these events in the next steps of this work. 

\noindent \textbf{(2) Flare (F): }
This category contains events whose light curves show a typical shape of stellar flare, 
i.e. with a quick rise and a gradual decay. 
Meanwhile, we require the events in this category to present at least two different levels of $\mathrm{CR}$, or at least one changepoint within the flaring phase. 
These events are most useful for our analyses, and 
their time parameters, peak luminosities and total energies are readily calculated. 

\noindent \textbf{(3) Flare with single Bayesian Block (S): }
This category contains events whose light curves show a clear rise and decay, 
but with only two major changepoints during the full exposure, 
i.e. there is only one single $\mathrm{CR}$ level of Bayesian Block within the flaring phase. 
The X-ray durations, peak luminosities and total energies are calculated accurately for these cases. 
However, the rise time and decay time cannot be determined accurately and set to be equal according to our definitions (see Section \ref{section_flare_properties_and_error_analysis}). 

\noindent \textbf{(4) Rise (R): }
These are events whose light curves display a quick rise, but do not fall back to the quiescent level before the exposure ends. 
That is to say, there is no changepoints after which 
$\Delta \mathrm{CR}$ satisfies $\Delta \mathrm{CR} \leq 5 \sigma_\mathrm{CR}$ consecutively. 
Because of the incomplete time coverage, these cases can only show the lower limits of their durations, peak luminosities and total energies. 
The decay times are not available in these cases. 
The rise times are measurable only when the peak is covered in the light curve. 
Otherwise, only the lower limits to the rise times are available. 

\noindent \textbf{(5) Decay (D): }
These are events which start from a level higher than the quiescent one but show a gradual decay. 
That is to say, there is no changepoints before which 
$\Delta \mathrm{CR}$ satisfies $\Delta \mathrm{CR} \leq 5 \sigma_\mathrm{CR}$ consecutively. 
Similarly to category R, these cases only show the lower limits of their durations, peak luminosities and total energies. 
The rise times are not available in these cases. 
The decay times are measurable only when the peak is covered in the light curve. 
Otherwise, only the lower limits to the decay times are available. 

\noindent \textbf{(6) Flare with multiple flaring phases (M): }
This category contains events whose light curves show two or more flaring phases. 
More specifically, each flaring phase has count levels with $\Delta \mathrm{CR} > 5 \sigma_\mathrm{CR}$; 
however, between two consecutive flaring phases, there exists at least one count level with $\Delta \mathrm{CR} \leq 5 \sigma_\mathrm{CR}$. 
In these cases, the $\mathrm{CR}$ reaches an earlier peak, falling back to the quiescent level, and it then goes up again. 
Such a process may be repeated several times. 
Therefore, we regard them as showing multiple flares within one CXO exposure. 
These cases are classified as multiple flare events, each of which can be classified into F, S, R or D categories 
according to the 
characteristics of the flaring phase. 
In the following statistics, we selectively treat each flaring phase as a single flare event depending on the parameters we analyse. 

\noindent \textbf{(7) Variable (V): }
Some events show strong variability 
(with some Bayesian Blocks with $\Delta \mathrm{CR} > 5 \sigma_\mathrm{CR}$ 
while others with $\Delta \mathrm{CR} \leq 5 \sigma_\mathrm{CR}$), 
but without a typical shape of quick rise and gradual decay. 
In a few cases, there are even quasi-periodic variations. 
These phenomena are possibly caused by transits between binaries or in multiple-body systems. 
Apparently, it is quite difficult for us to search for stellar flare events among these cases, 
because even though a star flares in such systems, 
the flare signals are mostly obscured by a non-flare variation. 
Thus, we exclude these events. 

\noindent \textbf{(8) Bad (B): }
There are mainly two reasons causing bad qualities of the light curves. 
One possible reason is that some sources are located near or at the edge of one single CCD. 
With the vibration of the instrument 
etc., not all of the photons coming from the source could be received, causing a dither effect \footnote{\url{https://cxc.harvard.edu/proposer/POG}}. 
Another possible reason is that the source is too faint to emit enough photons to the detector, 
as a result of distance or the effect of interstellar medium etc. 
In the above cases, the signal cannot be clearly delineated in the light curve. 
Events in this category could not be used for our analyses. 

\indent Among the eight categories as elucidated above, 
the categories C, F, R, D and V follow the criteria adopted by \citet{Getman2021ApJ916.32} with some modifications, 
while the other three categories S, M and B are newly defined for the sake of covering all types of light curves retrieved from our data. 

\begin{figure*}
	\centering
	\includegraphics[width=0.8\textwidth]{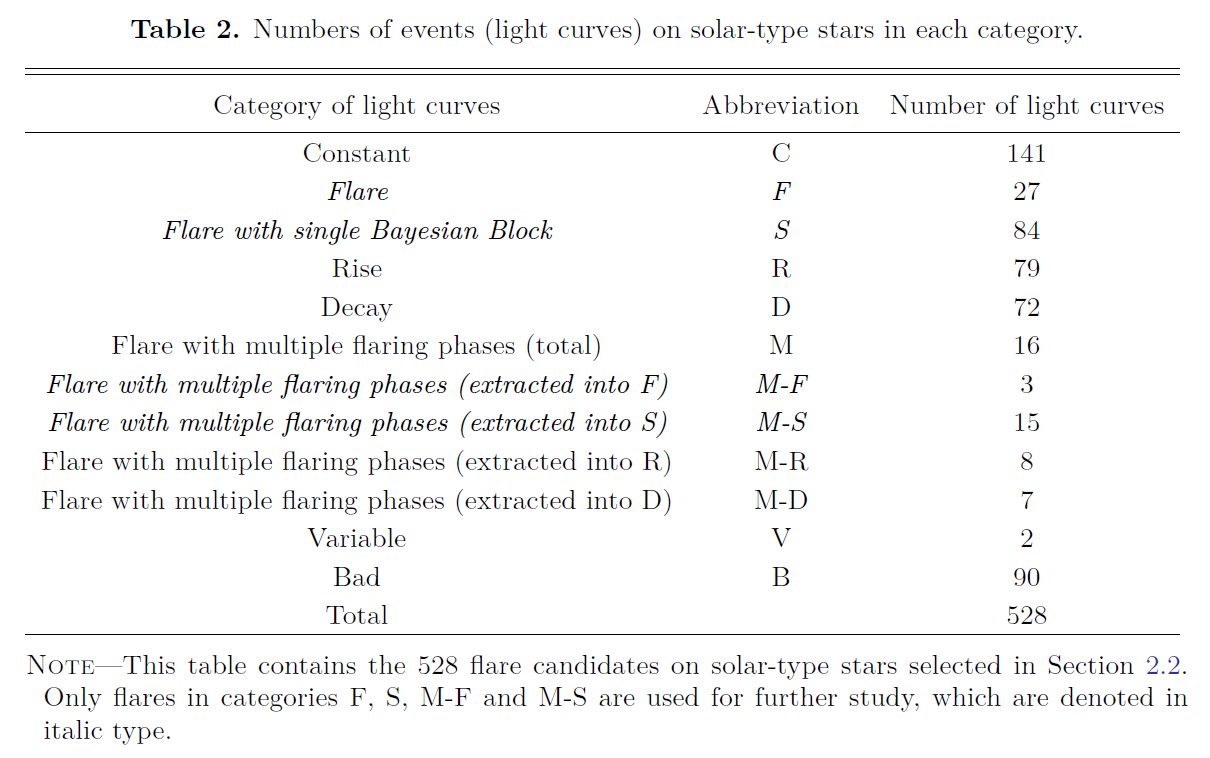}
	\caption{Table 2}
	\label{numbers_of_samples}
\end{figure*}

\indent Here, we use the events in categories F and S (including F and S events extracted from category M, denoted as M-F and M-S respectively) to measure the durations, peak luminosities and total energies, and the events in categories F and M-F to measure the rise and decay times. 
Among the 528 flare candidates on solar-type stars, 30 events belong to category F or M-F and 99 events belong to category S or M-S. 
The total number of individual sources in our sample set is 103. 
The numbers of events (light curves) in each category are listed in Figure \ref{numbers_of_samples}. 

\begin{figure} 
        \includegraphics[width=0.5\textwidth]{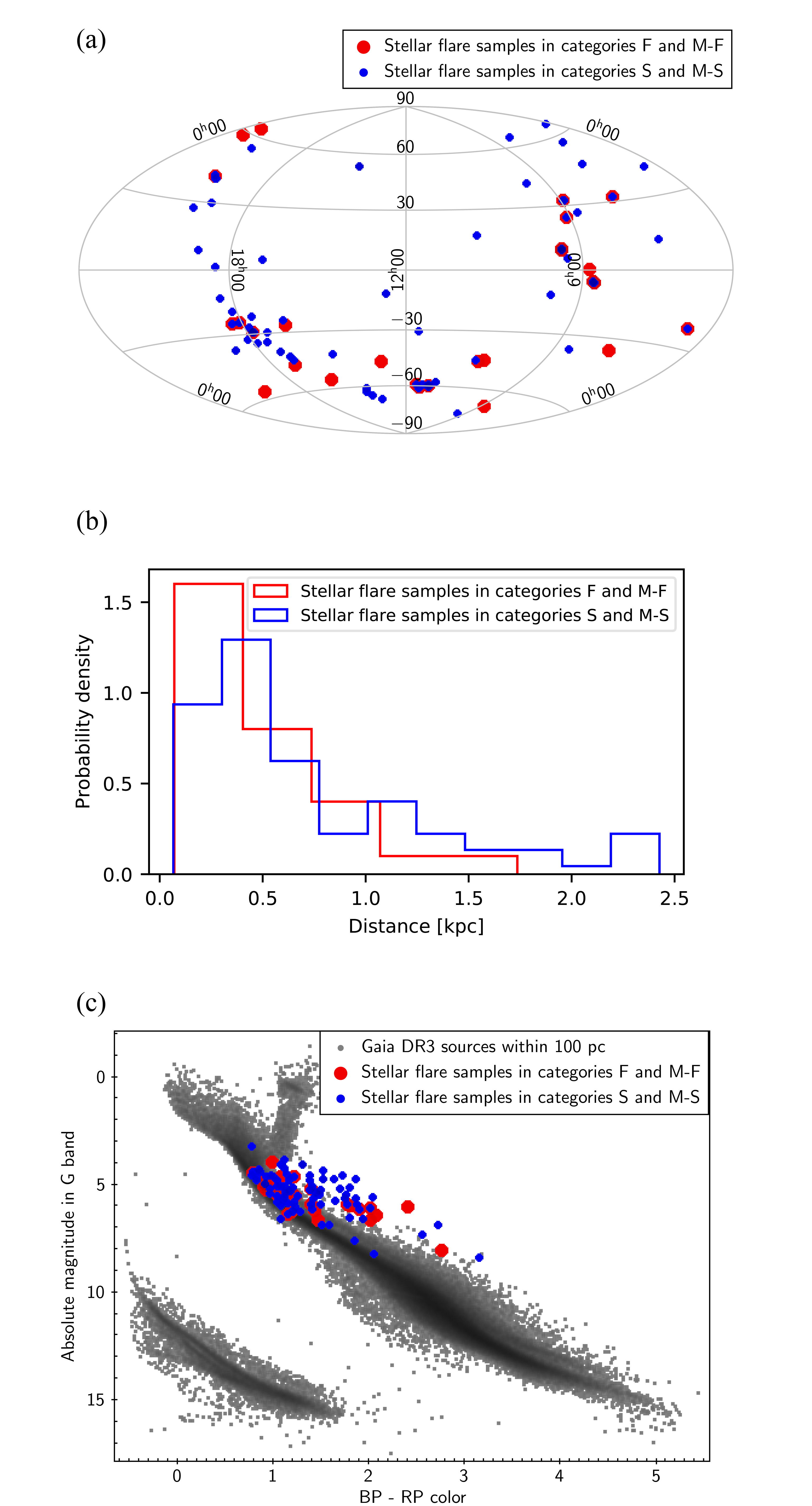}
	\caption{Distributions of the stellar flares in our sample with light curve categories F, M-F, S and M-S. 
 (a) Equatorial distribution. 
 (b) Distance distribution. 
 (c) Distribution across a Hertzsprung–Russell diagram (HRD). 
 When plotting the HRD, we use BP-RP colour 
 from Gaia DR3 as the x-axis, and use absolute magnitude in the G-band of Gaia (calculated from the G-band mean magnitude 
 and distance) from Gaia DR3 as the y-axis. 
 The probability density in panel (b) is defined as the value of the probability density function at each bin, which is normalized such that the integral over the range is $1$. Such a definition also applies for the probability density in the following figures. 
 }
\label{locations+distances+HRD}
\end{figure}

\indent The 103 flare host stars are cross-matched with the SIMBAD catalog\footnote{\url{http://simbad.u-strasbg.fr/simbad}} to check if there are any confirmed binary stars, and none of these sources are flagged as binary systems. 
However, there are still potential unresolved binary stars in our sample set, which is limited primarily by the photometric and spectrometric accuracies nowadays. 
The renormalised unit weight error (RUWE) from Gaia DR3 can serve as an indicator, since unresolved binary stars tend to show large RUWE values, e.g. $> 1.4$ \citep[e.g., ][]{Stassun2021ApJ.907.L33, Chulkov2022MNRAS517.2925}. 
The RUWE values of the flare-host stars are listed as a column of Figure \ref{properties_star}. 
Besides, those distribute above the mail-sequence band in Fig. \ref{locations+distances+HRD}c are especially likely to be binary systems, since they show higher luminosities than most main-sequence stars with the same temperatures. 

\indent The equatorial positions of the samples are illustrated in 
Fig. \ref{locations+distances+HRD}a; 
the probability distribution of the distances of the flare host stars is shown in 
Fig. \ref{locations+distances+HRD}b; 
and the samples in an HRD are plotted in 
Fig. \ref{locations+distances+HRD}c. 

\subsection{Flare properties and error analysis} \label{section_flare_properties_and_error_analysis}
\noindent The start time $t_\mathrm{start}$ and end time $t_\mathrm{end}$ are defined in Section \ref{section_analysis_of_light_curves}. 
The peak time $t_\mathrm{peak}$ of a flare is defined as the median time of the Bayesian Block with the highest $\mathrm{CR}$ during the flaring phase. 
Other useful time parameters of stellar flares, including the rise time $T_\mathrm{rise}$, decay time $T_\mathrm{decay}$ and total duration $T_\mathrm{duration}$, are defined as follows: 
\begin{eqnarray}
T_\mathrm{rise}     =& t_\mathrm{peak} - t_\mathrm{start}, \\
T_\mathrm{decay}    =& t_\mathrm{end}  - t_\mathrm{peak}, \\
T_\mathrm{duration} =& t_\mathrm{end}  - t_\mathrm{start}. 
\end{eqnarray}
Time parameters $t_\mathrm{start}$, $t_\mathrm{peak}$ and $t_\mathrm{end}$ are illustrated 
in Fig. \ref{lc_examples} 
with dash-dotted lines in green, red and blue, respectively. 

\indent In order to obtain flare parameters such as spectral indices, peak luminosities and total energies etc., 
we retrieve the spectra of the total 129 flare samples within each target region as mentioned in Section \ref{section_analysis_of_light_curves}. 
The spectra are extracted with the command \verb|specextract| in the software package CIAO\footnote{\url{https://asc.harvard.edu/ciao}} by the Chandra X-Ray Center, 
and from 5 different types of phases: quiescent phase, flaring phase, rise phase, decay phase and peak phase. 
The definitions of quiescent phase and flaring phase are already mentioned in Section \ref{section_analysis_of_light_curves}. 
The rise phase is defined as the time interval $(t_\mathrm{start},\ t_\mathrm{peak})$ for events in categories F or S, or $(t_{\mathrm{start},i},\ t_{\mathrm{peak},i})$ as the $i\mathrm{th}$ rise phase for events in categories M-F or M-S. Similarly, 
the decay phase is defined as the time interval $(t_\mathrm{peak},\ t_\mathrm{end})$ for events in categories F or S, or $(t_{\mathrm{peak},i},\ t_{\mathrm{end},i})$ as the $i\mathrm{th}$ decay phase for events in categories M-F or M-S. 
The peak phase is defined as the time interval of the Bayesian Block with the highest $\mathrm{CR}$ during each flaring phase, which could be denoted as $(t_\mathrm{peak,start},\ t_\mathrm{peak,end})$ for events in categories F or S, where $t_\mathrm{peak,start}$ and $t_\mathrm{peak,end}$ are the start and end time of this Bayesian Block respectively, 
or $(t_{\mathrm{peak,start},i},\ t_{\mathrm{peak,end},i})$ as the $i\mathrm{th}$ peak phase for events in categories M-F or M-S. 
We set the spectrum for the quiescent phase as the background of the other spectra with the command \verb|fparkey| in CIAO, thus the quiescent radiation of the star and the background radiation are deducted, leaving the pure radiation of the flare. 

\indent All of the retrieved spectra are then adaptively binned and fitted. 
The fitting procedure is performed with NASA's X-ray spectral fitting package Xspec\footnote{\url{https://heasarc.gsfc.nasa.gov/xanadu/xspec}}, 
and the model we use, i.e. the absorbed power-law spectra with pile-up, could be expressed as \verb|pileup*TBabs*powerlaw| in Xspec. 
The \verb|pileup| and \verb|TBabs| components are convolved into the model to consider the pile-up effect \citep[e.g., ][]{Davis2001ApJ562.575} and absorption from the interstellar medium respectively. 
An unabsorbed power-law spectrum is formulated as 
\begin{equation}
    \frac{\mathrm{d} N_\mathrm{photon}}{\mathrm{d} t \ \mathrm{d} E} = K E^{-\alpha}, 
\end{equation}
where $\frac{\mathrm{d} N_\mathrm{photon}}{\mathrm{d} t \ \mathrm{d} E}$ is in the units of $\mathrm{counts \ s^{-1} \ erg^{-1}}$, $K$ is the normalization coefficient and $\alpha$ is the dimensionless spectral index. 
After doing so, we obtain the spectral indices in the flare rise and decay phases, 
which are denoted as $\alpha_\mathrm{rise}$ and $\alpha_\mathrm{decay}$, respectively. 

\indent We obtain corrected fluxes within $0.5-8.0 \ \mathrm{keV}$ for the flaring phase (denoted as $F_\mathrm{flare}$) and peak phase (denoted as $F_\mathrm{peak}$) by using the \verb|flux| command to the \verb|powerlaw| component. 
And their errors are calculated from the normalization coefficients $K$, i.e. 
\begin{equation}
\sigma_{F}^{\pm} = F \frac{\sigma_K^{\pm}}{K}, 
\end{equation}
where $\sigma_F^{\pm}$ denotes the positive and negative errors of fluxes in linear scale. 
It's worthy of noticing that both $F$ and $\sigma_F^{\pm}$ are corrected by a factor of $\frac{1}{0.9}$ 
because the target region, i.e. the circle with the radius of $r_{90}$, only encircles $90\%$ of the total flux of the source (see Section \ref{section_analysis_of_light_curves}). 
Then the average luminosities for the flaring phase or peak phase could be obtained with 
\begin{equation} \label{L}
    L = 4 \pi d^2 F, \\
\end{equation}
assuming that the radiation from the star during the flaring phase is isotropic. 
The average luminosity $L_\mathrm{peak}$ during the peak phase is a good estimation of the real peak luminosity of a flare event when $t_\mathrm{peak,end} - t_\mathrm{peak,start}$ is relatively small. 
In such a case, $L_\mathrm{peak}$ is regarded as a proxy of the real peak luminosity. 
The errors of the luminosities are expressed as 
\begin{equation} \label{sigma_L}
    \sigma_L^{\pm} = \sqrt{\left(8 \pi d F\right)^2 {\sigma_d^{\pm}}^2 + \left( 4 \pi d^2 \right)^2 {\sigma_F^{\pm}}^2} 
\end{equation}
according to the error propagation formula, 
where $d$ and $\sigma_d$ are derived 
in Section \ref{section_selection_of_flare_candidates}. 
Equations (\ref{L}) and (\ref{sigma_L}) are calculated for both the whole flaring phase and peak phase. 
The average luminosities and their errors during the flaring phase are denoted as $L_\mathrm{flare}$ and 
$\sigma_{L_\mathrm{flare}}^{\pm}$ respectively, while the average luminosities and their errors during the peak phase are denoted as $L_\mathrm{peak}$ and $\sigma_{L_\mathrm{peak}}^{\pm}$, respectively. 

\indent With the average luminosity during the flaring phase $L_\mathrm{flare}$, we then estimate the total energy released by the flare with 
\begin{equation} \label{E=LT}
    E_\mathrm{flare} = \int_{t_\mathrm{start}}^{t_\mathrm{end}} L(t) \ \mathrm{d} t = 
    L_\mathrm{flare} T_\mathrm{duration}, 
\end{equation}
and the error of $E_\mathrm{flare}$ is 
\begin{equation}
    \sigma_{E_\mathrm{flare}}^{\pm} = \sigma_{L_\mathrm{flare}}^{\pm} T_\mathrm{duration}. 
\end{equation}
Note that, $\sigma_{L_\mathrm{peak}}^{\pm}$ and $\sigma_{E_\mathrm{flare}}^{\pm}$ are estimated by a combination of the error in distance and that arising from the \verb|powerlaw| model in Xspec while fitting the spectra.

\subsection{Estimation of flare frequencies} \label{section_estimation_of_flare_frequencies}
\noindent In order to obtain the frequency distribution of flares, we inspect a bigger data set which contains all of the solar-type stars both with and without detected flare events, and use the same rules to select our sample set as listed in Figure \ref{numbers_of_samples}. 
More specifically, we identify such a data set with the same criteria mentioned in Sections \ref{section_selection_of_flare_candidates} and \ref{section_selection_of_solar-type_stars}, but without checking Variability Indices and light curves, 
because all of the other selection rules are unrelated to stellar activities. 
The total number of individual detections (per source per CXO exposure) in this data set is 8,016, with 2,748 individual sources and 2,653 independent CXO exposures. 
We denote the effective exposure time of each individual detection as $\tau_n$, which is obtained from the total elapsed time of the valid data of each observation, 
\verb|gti_elapse|, available in CSC 2.0 data\footnote{\url{https://cxc.harvard.edu/csc/columns/persrc.html}}. 

\indent We use the following equation to estimate the flare frequency distribution 
as a function of flare energy on solar-type stars:  
\begin{equation} \label{FDF_E}
    f(E_\mathrm{flare}) \ \mathrm{d} E_\mathrm{flare} = 
    \frac{\Delta N_\mathrm{flare} (E_\mathrm{flare})}
         {\tau \ \Delta E_\mathrm{flare}} \ \mathrm{d} E_\mathrm{flare}, 
\end{equation}
where $\Delta N_\mathrm{flare}$ is the number of detected flares in each flare energy bin, specifically meaning the number of events in light curve categories F, S, R, D, M-F, M-S, M-R and M-D. 
Considering that it is impossible to accurately calculate the flare energies of flares in categories R, D, M-R and M-D, we omit these four categories, which means that the flare frequency $f(E_\mathrm{flare})$ in equation (\ref{FDF_E}) is only a lower limit. 
In equation (\ref{FDF_E}), 
$\Delta E_\mathrm{flare}$ represents the energy range of each flare energy bin, and 
$\tau$ is the total effective exposure time of all of the 8,016 individual CXO detections and is calculated by 
\begin{equation}
    \tau = \sum_{n=1}^{8016} \tau_n = 9.98 \ \mathrm{yr}. 
\end{equation}
We can see that such a flare frequency distribution function $f(E_\mathrm{flare})$ is in the units of 
$\mathrm{flares} \ \mathrm{star}^{-1} \ \mathrm{yr}^{-1} \ \mathrm{erg}^{-1}$. 

\indent After defining the frequency distribution function, 
we further define the corresponding normalised cumulative occurrence probability as a function of flare energy: 
\begin{equation} \label{CFDF_E} 
    \tilde{f}(>E_\mathrm{flare}) = \frac
    {\int_{E_\mathrm{flare}}^{+\infty} f(E_\mathrm{flare}) \ \mathrm{d} E_\mathrm{flare}}
    {\int_0^{+\infty} f(E_\mathrm{flare}) \ \mathrm{d} E_\mathrm{flare}}. 
\end{equation}
Similarly, this equation only gives a lower limit to the actual cumulative distribution, 
since only F, S, M-F and M-S samples have been taken into account.

\section{Results and discussions} \label{section_results_and_discussions}
\noindent After determining the physical quantities about stellar flares and corresponding host stars by using the methods in Section \ref{section_data_and_methods}, we present the statistical results in this Section. 
Physical quantities in all figures are denoted with a subscript SXR, in order to specify that these quantities refer to the CXO soft X-ray band. 

\indent The probability density functions of some quantities in logarithmic scale, such as time parameters, peak luminosities, flare energies and spectral indices,  can be well fitted by normal distributions. 
Note that, we apply normal distributions here only to statistically estimate the corresponding mean values and standard deviations of the key parameters without specific physical assumptions. 

\begin{figure*}
	\centering
	\includegraphics[width=1.0\textwidth]{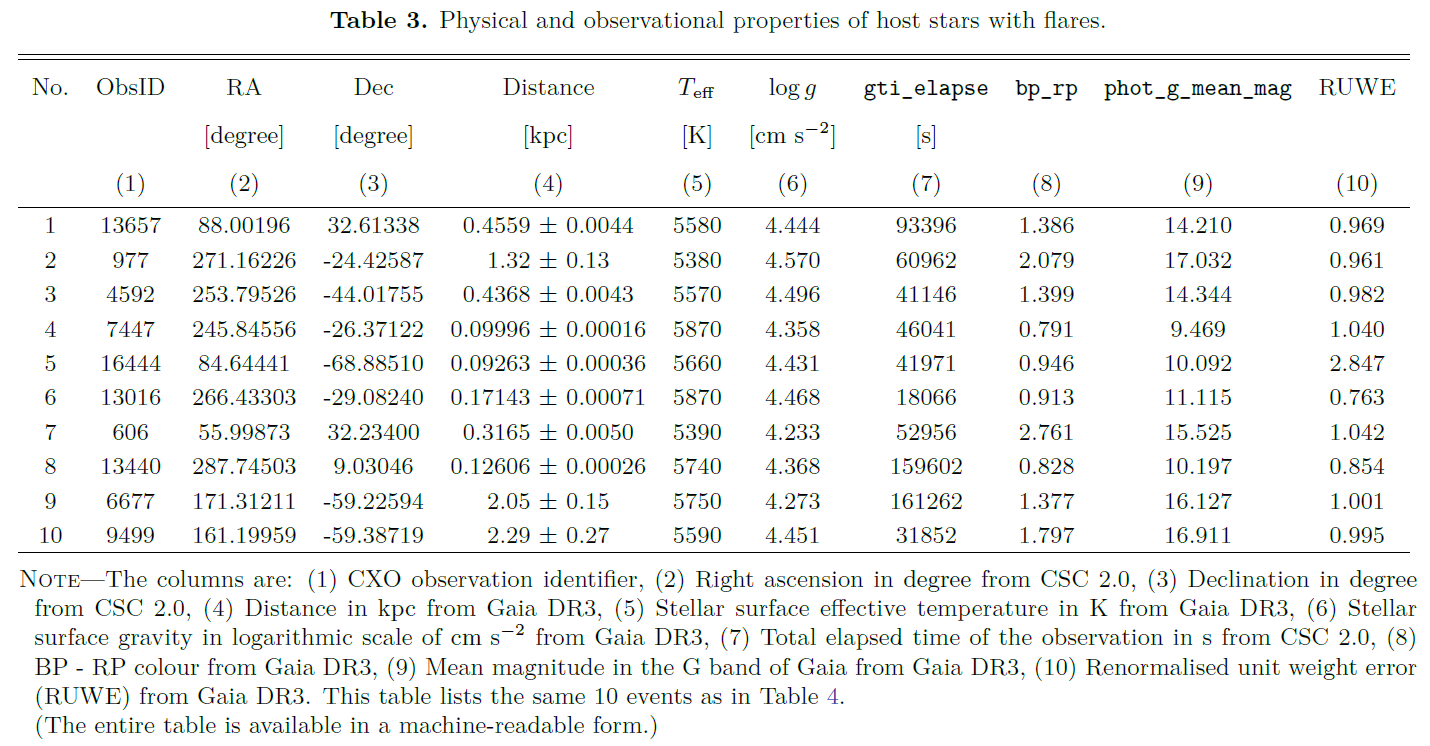}
	\caption{Table 3}
	\label{properties_star}
\end{figure*}
\begin{figure*}
	\centering
	\includegraphics[width=1.0\textwidth]{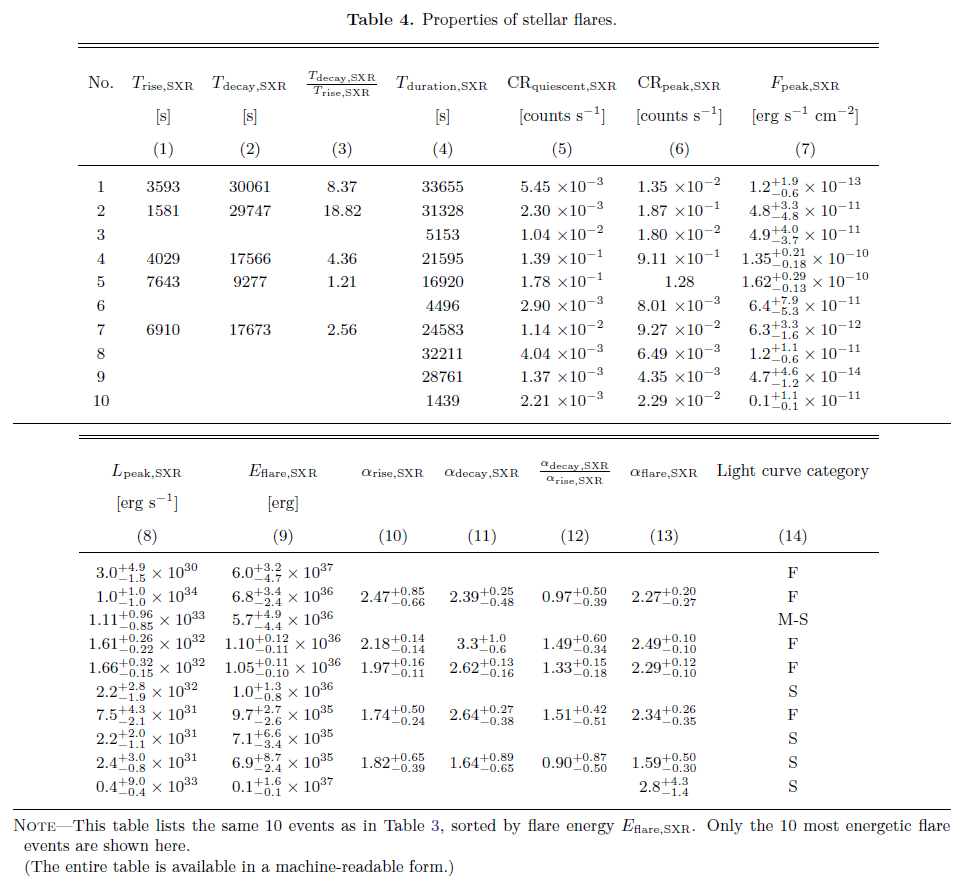}
	\caption{Table 4}
	\label{properties_flare}
\end{figure*}

\indent Physical and observational properties of flare host stars are listed in Figure \ref{properties_star}, and the properties of corresponding stellar flare events are listed in Figure \ref{properties_flare}. 
Only 10 events with the highest flare energies are shown in the tables as examples, while the entire tables are available online in a machine-readable form. 

\subsection{Distributions of time parameters}
\begin{figure}
    \centering
    \includegraphics[width=0.5\textwidth]{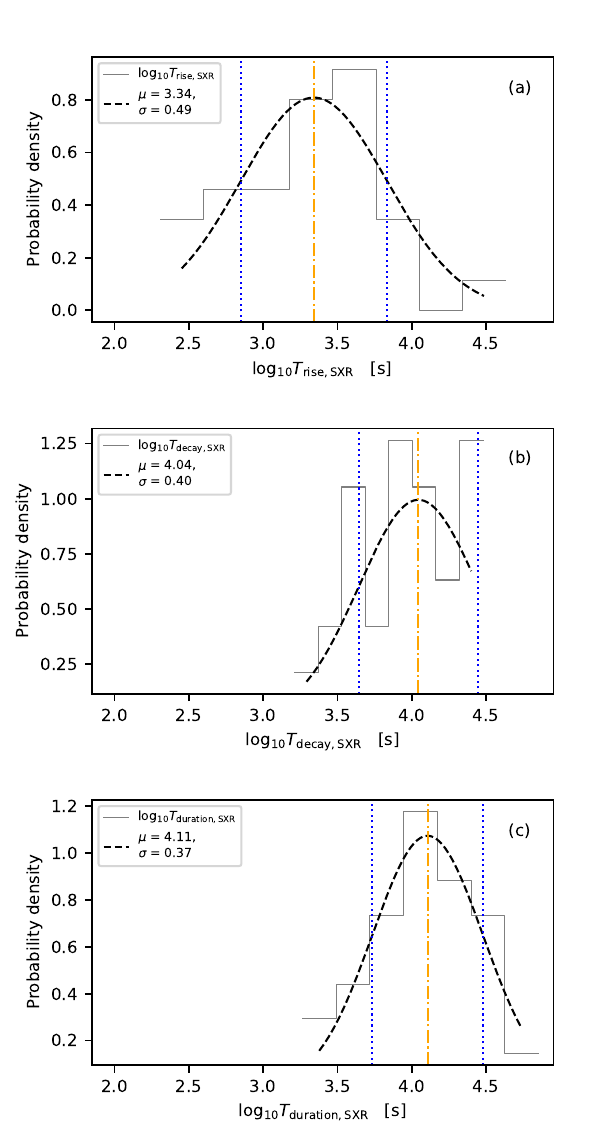}
    \caption{Distributions of time parameters for F and M-F events. (a)-(c) Histograms and corresponding normal distributions of $T_\mathrm{rise,SXR}$, $T_\mathrm{decay,SXR}$ and $T_\mathrm{duration,SXR}$ in logarithmic scale. 
    The most probable values ($\mu$) and standard deviations ($\sigma$) are listed in the panel legends. 
    The values of $\mu$ are illustrated with orange dash-dotted lines, and the values of $\mu-\sigma$ and $\mu+\sigma$ are illustrated with blue dotted lines in each panel.}
    \label{distribution_tp_FMF}
\end{figure}
\begin{figure}
    \centering
    \includegraphics[width=0.5\textwidth]{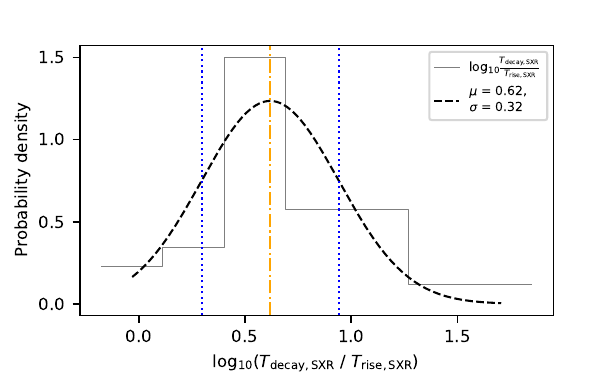}
    \caption{Distribution of the ratio of decay time to rise time for F and M-F samples. 
    Panel legend and the meaning of line styles are the same as in Fig. \ref{distribution_tp_FMF}, but for $\log_{10} (T_\mathrm{decay,SXR} / T_\mathrm{rise,SXR})$. }
    \label{distribution_T_ratio}
\end{figure}

\begin{figure}
    \centering
    \includegraphics[width=0.5\textwidth]{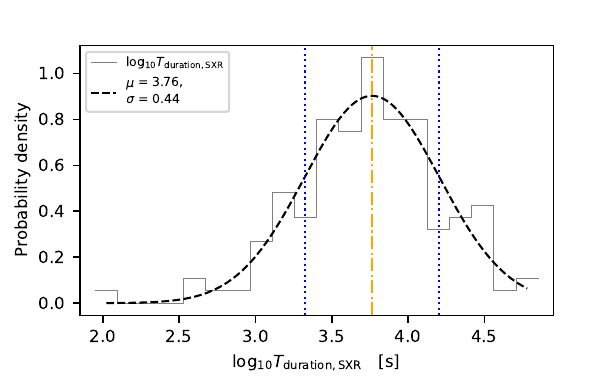}
    \caption{Distribution of flare durations for all F, M-F, S and M-S events. 
    Panel legend and the meaning of line styles are the same as in Fig. \ref{distribution_tp_FMF}. }
    \label{distribution_T_duration_FMFSMS}
\end{figure}

\noindent We obtain the time parameters of flares in categories F, M-F, S and M-S. 
The flare rise time $T_\mathrm{rise,SXR}$ and the decay time $T_\mathrm{decay,SXR}$ are only measured for 30 F and M-F samples, 
while the flare duration $T_\mathrm{duration,SXR}$ is measured for all of the 129 F, M-F, S and M-S samples. 

\indent Distributions of time parameters $T_\mathrm{rise,SXR}$, $T_\mathrm{decay,SXR}$ and $T_\mathrm{duration,SXR}$ of F and M-F samples are shown in 
Fig. \ref{distribution_tp_FMF}. 
The flare rise time $T_\mathrm{rise,SXR}$ spans about two orders of magnitude, ranging from $\sim 10^{2.5} \ \mathrm{s}$ to $\sim 10^{4.5} \ \mathrm{s}$ 
and centering at $10^{3.34 \pm 0.49(1\sigma)} \ \mathrm{s}$. 
The flare decay time $T_\mathrm{decay,SXR}$ spans about one order of magnitude, ranging from $\sim 10^{3.5} \ \mathrm{s}$ to $\sim 10^{4.5} \ \mathrm{s}$ 
and centering at $10^{4.04 \pm 0.40(1\sigma)} \ \mathrm{s}$. 
The flare duration $T_\mathrm{duration,SXR}$ spans about one order of magnitude, ranging from $\sim 10^{3.5} \ \mathrm{s}$ to $\sim 10^{4.5} \ \mathrm{s}$ 
and centering at $10^{4.11 \pm 0.37(1\sigma)} \ \mathrm{s}$. 
It is found that the decay phase occupies most of the entire flare duration in most of events.

\indent Distribution of the ratio of decay time to rise time $T_\mathrm{decay,SXR} / T_\mathrm{rise,SXR}$ for F and M-F samples is shown in Fig. \ref{distribution_T_ratio}. 
The ratio of decay time to rise time ranges from $\sim 1$ to $\sim 10^{1.5}$ with an average of $10^{0.62 \pm 0.32(1\sigma)}$. 
The majority of flares have $T_\mathrm{decay,SXR} > T_\mathrm{rise,SXR}$, 
except for only one event of F or M-F category showing $T_\mathrm{decay,SXR} / T_\mathrm{rise,SXR} = 0.668$. 
However, we suspect that this special event is a result of two or more overlapping events that are not strictly identified by our identification methods (see Section \ref{section_analysis_of_light_curves}), as shown in Fig. \ref{obsid_1872} in Appendix \ref{appendix_a_special_light_curve}. 
We thus conclude that the SXR profile of stellar flares consists of a rise phase and a much longer decay phase, very similar to the characteristic of 
solar SXR flares, 
even though the stellar flares show a much longer duration than the solar flares, 
which typically last for $\sim 10^{2}$ to $\sim 10^{4} \ \mathrm{s}$ with a median duration of $\sim 400 \ \mathrm{s}$ \citep{Veronig2002A&A382.1070}. 
For solar flares, it was found that the association between flares and CMEs increases with the class and/or duration of flares \citep[e.g. ][]{Cheng2010ApJ712.752}. 
Thus the flares in F or M-F category, in particular for the longest-duration ones, are most likely accompanied by stellar CMEs. 

\indent The distribution of flare duration for all F, M-F, S and M-S samples is displayed in Fig. \ref{distribution_T_duration_FMFSMS}. 
The flare duration is found to range from $\sim 10^{2.0} \ \mathrm{s}$ to $\sim 10^{4.5} \ \mathrm{s}$ with a median of $10^{3.76 \pm 0.44(1\sigma)} \ \mathrm{s}$. 
It is obvious that, after including S and M-S samples, the distribution of $T_\mathrm{duration,SXR}$ moves towards the shorter time scale. 
It indicates that S and M-S flares last for a shorter time than F and M-F flares. 
The difference between the two groups is mainly due to two reasons: 
(1) the events with a shorter duration tend to be described with fewer blocks when fitted with the Bayesian Block method, 
or (2) the duration of S and M-S events can be underestimated when the corresponding changepoints are regarded as $t_\mathrm{start}$ and $t_\mathrm{end}$ as mentioned in Sections {\ref{section_analysis_of_light_curves}} and {\ref{section_flare_properties_and_error_analysis}}. 

\subsection{Distributions of peak luminosities and flare energies} \label{section_distributions_of_peak_luminosities_and_flare_energies}
\begin{figure*}
    \centering
    \includegraphics[width=1.0\textwidth]{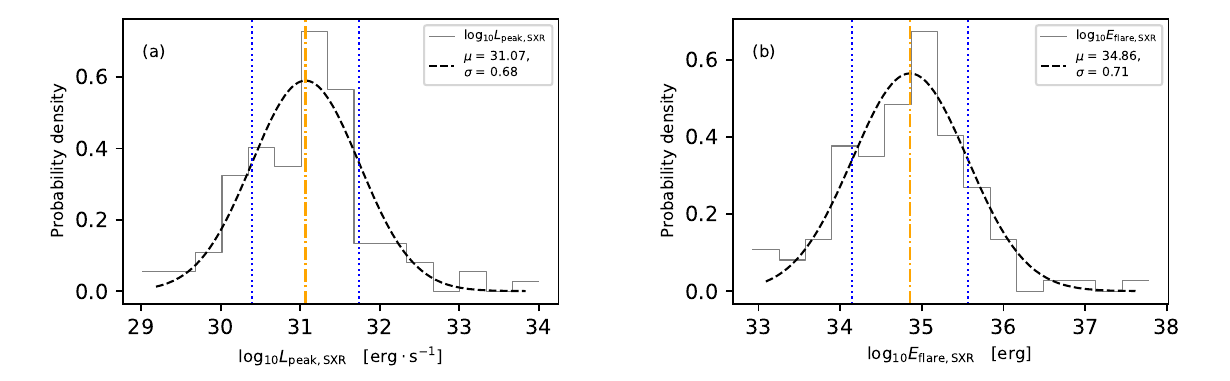}
    \caption{Distributions of peak luminosities (a) and flare energies (b). Panel legends and the meaning of line styles are the same as in Fig. \ref{distribution_tp_FMF}.}
    \label{distribution_L+E}
\end{figure*}
\noindent We obtain the flare peak luminosities and flare energies of 129 flare events in categories F, M-F, S and M-S. 

\indent Distributions of flare peak luminosities $L_\mathrm{peak,SXR}$ and flare energies $E_\mathrm{flare,SXR}$ are shown in Fig. \ref{distribution_L+E}a and \ref{distribution_L+E}b, respectively. 
The flare peak luminosity ranges from $\sim 10^{29} \ \mathrm{erg \ s^{-1}}$ to $\sim 10^{34} \ \mathrm{erg \ s^{-1}}$, 
centering at $10^{31.07 \pm 0.68(1\sigma)} \ \mathrm{erg \ s^{-1}}$. 
The flare energy varies from $\sim 10^{33} \ \mathrm{erg}$ to $\sim 10^{37} \ \mathrm{erg}$ with an average of $10^{34.86 \pm 0.71(1\sigma)} \ \mathrm{erg}$. 


\indent The energy of the strongest flare event among our samples is $6.0_{-4.7}^{+3.2} \times 10^{37} \ \mathrm{erg}$. 
It is detected in $\mathrm{ObsID=13657}$ and has $T_\mathrm{rise} = 3593 \ \mathrm{s}$ and $T_\mathrm{decay} = 30061 \ \mathrm{s}$. 
The corresponding host star is $0.4559 \pm 0.0044 \ \mathrm{kpc}$ from us and has a surface effective temperature of $5580 \ \mathrm{K}$ and surface gravity of $\log g = 4.444$. 
This host star is classified as a pulsating variable star according to \citet{Hartman2008ApJ675.1254}, after cross-matching with SIMBAD. 

\indent One of the most energetic solar flare event in record is known as the Carrington event, 
which occurred on September 1st 1859 \citep{Carrington1859MNRAS20.13} and 
released a total bolometric energy of $E_\mathrm{flare,bol} \sim 5 \times 10^{32} \ \mathrm{erg}$ \citep{Cliver2013JSWSC3.A31}. 
By comparison, the weakest flare from solar-type stars in our sample released an energy of $E_\mathrm{flare,SXR} = 0.9_{-0.4}^{+5.0} \times 10^{33} \ \mathrm{erg}$. 
It is worth mentioning that the flare energy emitted at the SXR band is only a fraction ($\sim 1/10 - 1/4$) of that at the optical band 
for events with $E_\mathrm{flare,SXR}$ between $\sim 10^{34}$ and $\sim 10^{36} \ \mathrm{erg}$ \citep{Flaccomio2018AA620.A55} on PMS stars. 
\citet{Kretzschmar2011A&A530.A84} also suggested that, for solar flares, 
the energy at the optical band represents about two thirds of the bolometric energy. 
Thus, if the flares on stars originate from the similar physical mechanism to those on the Sun and show a similar energy fraction at the SXR band, 
the energy of flares at the SXR band $E_\mathrm{flare,SXR}$ may constitute only a minor fraction of the total bolometric energy $E_\mathrm{flare,bol}$. 
Specifically, the calculated $E_\mathrm{flare,SXR}$ here is roughly $\sim 1/10$ of the bolometric energies $E_\mathrm{flare,bol}$. 

\begin{figure*}
	\centering
	\includegraphics[width=1.0\textwidth]{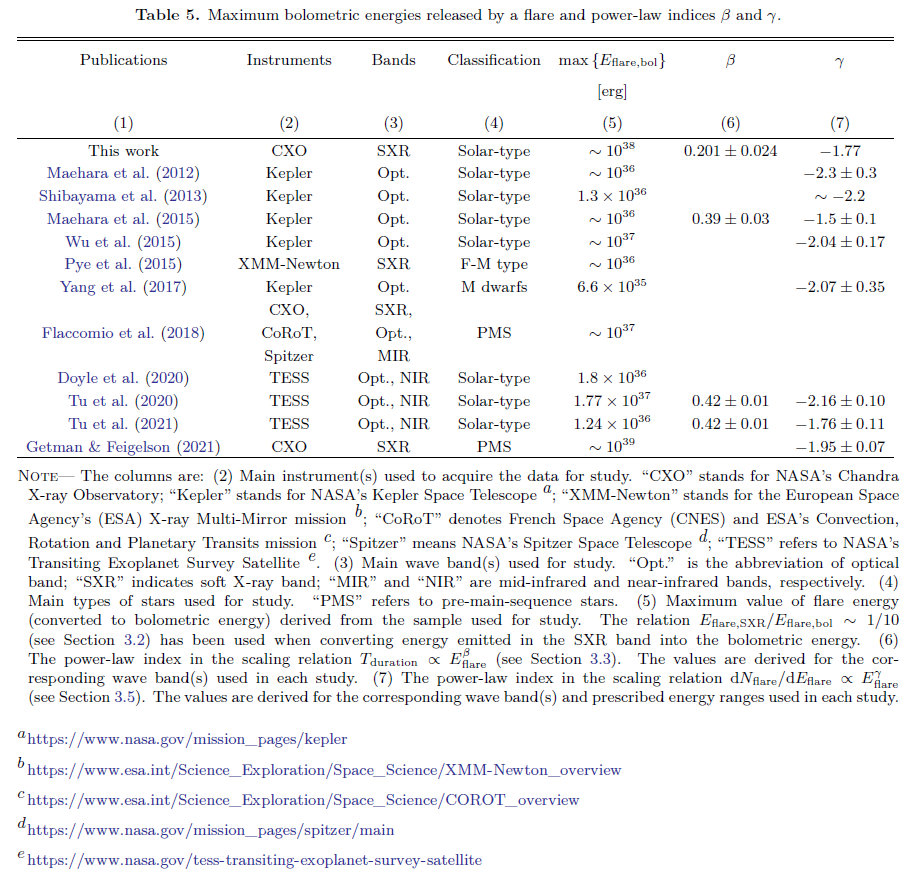}
	\caption{Table 5}
	\label{max_E+beta+gamma}
\end{figure*}

\indent By comparison, the maximum bolometric energies in previous studies are listed in column (5) of 
Figure \ref{max_E+beta+gamma}. 
We can see that in most cases, the maximum flare bolometric energy is $10^{36}-10^{38} \ \mathrm{erg}$ under the assumption that flares on different stars show a similar energy fraction at the SXR band, 
indicating that it is saturated at $\sim 10^{38}-10^{39} \ \mathrm{erg}$ above which nearly no flares have been detected. 
The upper limit of flare bolometric energy derived by \citet{Wu2015ApJ798.92} is $2 \times 10^{37} \ \mathrm{erg}$, which, however, might be underestimated 
as a result of selection effect of instruments or different methods of calculating flare energy. 
In this work, 
the flare energy is calculated by fitting the spectrum with an absorbed power-law spectrum and then 
calculating the average unabsorbed flux during the entire flaring phase (see Section \ref{section_flare_properties_and_error_analysis}), 
while \citet{Wu2015ApJ798.92} calculated the flare energy at the optical or NIR bands by assuming white-light flares as blackbody emission 
\citep[e.g., ][]{Maehara2012Nat485.478,Maehara2015EPS67.59,Shibayama2013ApJS209.5}. 

\indent Superflares are defined as flares that emit bolometric energies higher than the Carrington event, 
i.e. $E_\mathrm{flare,bol} \gtrsim 10^{33} \ \mathrm{erg}$ 
\citep[e.g., ][]{Maehara2012Nat485.478,Maehara2015EPS67.59,Doyle2020MNRAS494.3596,Tu2020ApJ890.46,Tu2021ApJS253.35}. 
Thus, the majority of stellar flares in our sample are superflares. 
\citet{Getman2021ApJ916.32} defined ``megaflares'' as events with flare energy in the SXR band $E_\mathrm{flare,SXR}$ exceeding $10^{36.2} \ \mathrm{erg}$. 
According to this definition, 3 out of the 129 flares 
in this work (i.e. No. $1-3$ in Figures \ref{properties_star} and \ref{properties_flare}) belong to ``megaflares''. 

\begin{figure}
	\centering
	\includegraphics[width=0.5\textwidth]{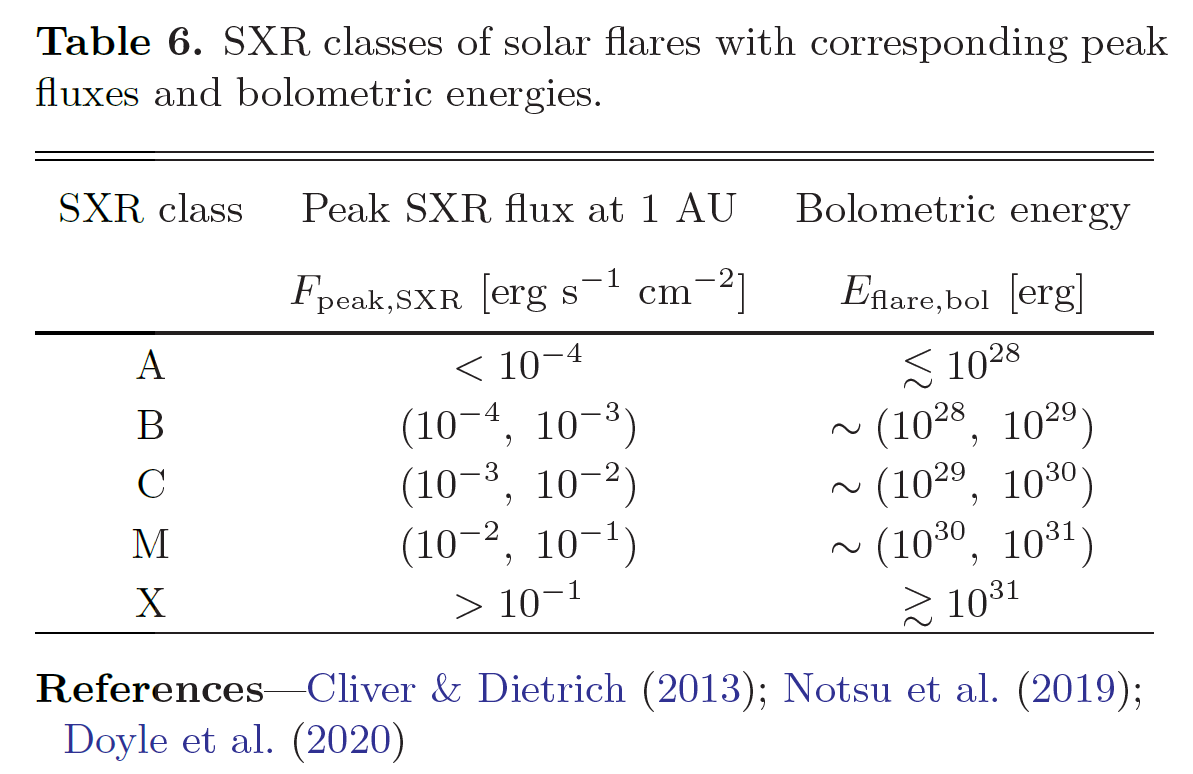}
	\caption{Table 6}
	\label{solar_flare_levels}
\end{figure}

\indent Moreover, solar flares are usually classified in terms of their Geostationary
Operational Environmental Satellite (GOES) $1-8 \ \text{\AA}$ peak SXR flux $F_\mathrm{peak,SXR}$. 
Figure \ref{solar_flare_levels} lists the flare classes and the corresponding GOES peak SXR flux \citep{Cliver2013JSWSC3.A31}, as well as the 
derived flare bolometric energy $E_\mathrm{flare,bol}$ \citep{Notsu2019AAS234.122.02,Doyle2020MNRAS494.3596}. 
The Carrington event was classified as an $\mathrm{X45 \ (\pm 5)}$ class flare according to \citet{Cliver2013JSWSC3.A31}. 
Since both GOES and CXO work at SXR bands, with the derived $F_\mathrm{peak,SXR}$ at $1 \ \mathrm{AU}$, we estimate that 
all of the flare events in this work should be $\mathrm{X}$-class ones, roughly ranging from $\mathrm{X100}$ to $\mathrm{X10,000,000}$, 
which are significantly greater than solar flares. 
Based on the fact that the Carrington event caused an enormous geomagnetic storm \citep{Tsurutani2003JGRA108.1268}, 
thus threatening communication or navigation systems \citep{Maehara2015EPS67.59}, 
it is reasonable to infer that the megaflares, at least superflares, as found here could have serious influences on the space weather and atmosphere of associated exoplanets if they exist. 

\subsection{Correlations among $T_\mathrm{duration}$, $L_\mathrm{peak}$ and $E_\mathrm{flare}$} \label{section_correlations_beteeen_TLE}
\begin{figure}
    \centering
    \includegraphics[width=0.5\textwidth]{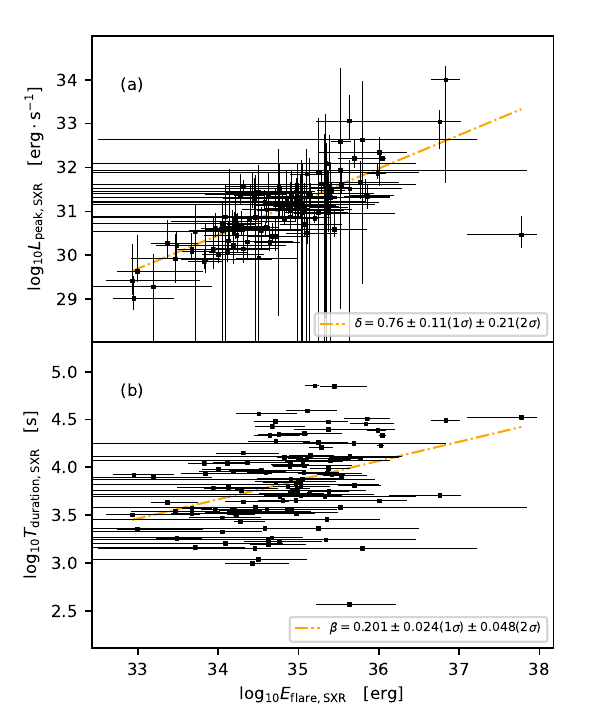}
    \caption{Scatter plots of (a) peak luminosity and (b) duration as a function of flare energy. 
    Error bars are plotted for each flare. 
    Errors of $L_\mathrm{peak,SXR}$ and $E_\mathrm{flare,SXR}$ are estimated from the uncertainty of distance and 
    the uncertainty of flux derived from the model when fitting the spectra (see Section \ref{section_flare_properties_and_error_analysis}). 
    Orange dash-dotted lines represent the best-fitting results of the power-law correlations 
    $L_\mathrm{peak,SXR}     \propto E_\mathrm{flare,SXR}^{0.76\pm0.11(1\sigma_\delta)}$ and 
    $T_\mathrm{duration,SXR} \propto E_\mathrm{flare,SXR}^{0.201\pm0.024(1\sigma_\beta)}$, 
    and the power-law indices are also shown in the legends of corresponding panels. 
    }
    \label{correlation_LTE}
\end{figure}

\begin{figure}
    \centering
    \includegraphics[width=0.5\textwidth]{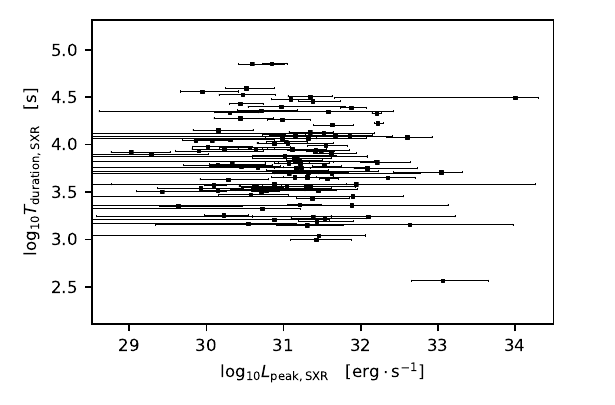}
    \caption{Scatter plot of the duration as a function of peak luminosity. 
    Error bars are plotted for each flare, as in Fig. \ref{correlation_LTE}. 
    }
    \label{correlation_TL}
\end{figure}

\noindent We further study the correlations between the time and energetic parameters of the flares, which are also fitted with the following power-law formulae: 
\begin{eqnarray}
L_\mathrm{peak,SXR}     &\propto E_\mathrm{flare,SXR}^{\delta}, \label{L_propto_E^delta}\\
T_\mathrm{duration,SXR} &\propto E_\mathrm{flare,SXR}^{\beta}. 
\end{eqnarray}

\indent The scatter plot of the peak luminosity against the flare energy is shown in Fig. \ref{correlation_LTE}a, and 
the best-fitting with a power-law relation is illustrated. 
We use $\sigma_\delta$ to denote the standard deviation of $\delta$, as obtained via Monte-Carlo simulations. 
Only the values whose lower limits are greater than zero are taken into account when doing the Monte-Carlo simulations. 
The power-law index is $\delta = 0.76 \pm 0.11(1\sigma_\delta)$. 
The Pearson correlation coefficient is $0.77$ (with a p-value of $5.6 \times 10^{-23}$), 
indicating that there is a strong correlation between $L_\mathrm{peak,SXR}$ and $E_\mathrm{flare,SXR}$. 

\indent From the power-law correlation of equation (\ref{L_propto_E^delta}), we can easily get 
\begin{equation}
E_\mathrm{flare,SXR} \propto L_\mathrm{peak,SXR}^{1 / \delta} 
\end{equation}
with $1 / \delta = 1.31 \pm 0.18(1\sigma)$, which is in good agreement with the scaling relation of 
$E_\mathrm{flare,SXR} \propto L_\mathrm{peak,SXR}^{\sim 1.2}$ derived by \citet{Pye2015AA581.A28} from their XMM-Newton survey of 
$\sim 130$ flares from $\sim 70$ Tycho cool (F-M type) stars. 
The similar scaling indices indicate that F-M type stars, including solar-type stars, 
may share the same mechanisms responsible for the energy relase at the SXR band. 

\indent A similar scaling index of 1.16 was also reported by \citet{Wolk2005ApJS160.423} from their Chandra survey of 41 K-type PMS stars, 
although they used the average luminosity during the entire flaring phase $(L_\mathrm{flare,SXR})$ rather than the peak luminosity as we use here. 
The similarity in the scaling index between $E_\mathrm{flare,SXR}$ and $L_\mathrm{peak,SXR}$ and that between $E_\mathrm{flare, SXR}$ and $L_\mathrm{flare,SXR}$ indicates that $L_\mathrm{flare,SXR}$ is nearly proportional to $L_\mathrm{peak,SXR}$. 

\indent Since $1/\delta$ is close to 1, we can say that the flare energy varies nearly linearly with the peak luminosity, which implies that the flare duration may be independent of its peak luminosity, 
as argued by \citet{Wolk2005ApJS160.423} and \citet{Pye2015AA581.A28}. 
To test this point, we make a scatter plot of $T_\mathrm{duration}$ against $L_\mathrm{peak,SXR}$ as shown in Fig. \ref{correlation_TL}, 
and obtain a Pearson correlation coefficient of $-0.095$ (with a p-value of $0.32$) between them, 
which confirms the independence between the two parameters. 

The scatter plot of the duration as a function of flare energy is shown in Fig. \ref{correlation_LTE}b. 
The power-law index is $\beta = 0.201 \pm 0.024(1\sigma_\beta)$, where $\sigma_\beta$ stands for the standard deviation of $\beta$, similar to $\sigma_\delta$ as mentioned above. 
The Pearson correlation coefficient is $0.38$ (with a p-value of $2.8 \times 10^{-5}$), 
indicating that there is a weak correlation between $T_\mathrm{duration,SXR}$ and $E_\mathrm{flare,SXR}$. 

\indent For comparison, the best-fitting index $\beta$ of $T_\mathrm{duration}$ as a function of $E_\mathrm{flare}$ in logarithmic scale from previous studies are listed in column (6) of Figure \ref{max_E+beta+gamma}. 
According to the magnetic reconnection theory of stellar flares, the flare energy originates from the release of magnetic energy and the duration is comparable to the reconnection time \citep[e.g., ][]{Shibata2016IAUS320.3}. 
Assuming that all solar-type stars have similar stellar magnetic properties (e.g., magnetic field strength $B$) and applying dimensional analysis, 
\citet{Maehara2015EPS67.59} derived a power-law index of $\beta \sim 1/3$ in the relationship 
$T_\mathrm{duration,bol} \propto E_\mathrm{flare,bol}^{\beta}$ 
, which is in good agreement with the results for solar flares \citep[e.g., ][]{Namekata2017ApJ851.91}. 
It is found that the indices derived by using the data at optical and NIR bands are usually consistent with the theoretical value of $\beta \sim 1/3$, 
such as $0.39 \pm 0.03$ obtained by \citet{Maehara2015EPS67.59} and $0.42 \pm 0.01$ by \citet{Tu2020ApJ890.46,Tu2021ApJS253.35}. 

\indent It is obvious that the value of $\beta$ ($0.201 \pm 0.024(1\sigma_\beta)$) derived here is much smaller than previous values. 
We calculate the Bayesian Information Criteria (BIC) values of our fit and the fit where the slope is fixed to $1/3$ and find that they are $-211.46$ and $-208.02$, respectively, with a difference of $3.44$. This reconfirms that the value of $\beta$ we derived is indeed significantly different from $\beta = 1/3$. 
Unlike optical (white-light) flares, which well represent the overall bolometric feature of flare events, 
the duration of the SXR flares increases more slowly with the energy than the bolometric ones. 
Such a significant difference indicates that 
the SXR emissions of stellar flares originate from different radiation mechanisms from the optical ones, 
as also reflected from the distinct energy partitions at the SXR and optical bands as derived in Section \ref{section_distributions_of_peak_luminosities_and_flare_energies}.

\subsection{Distributions of spectral indices}
\begin{figure}
    \centering
    \includegraphics[width=0.5\textwidth]{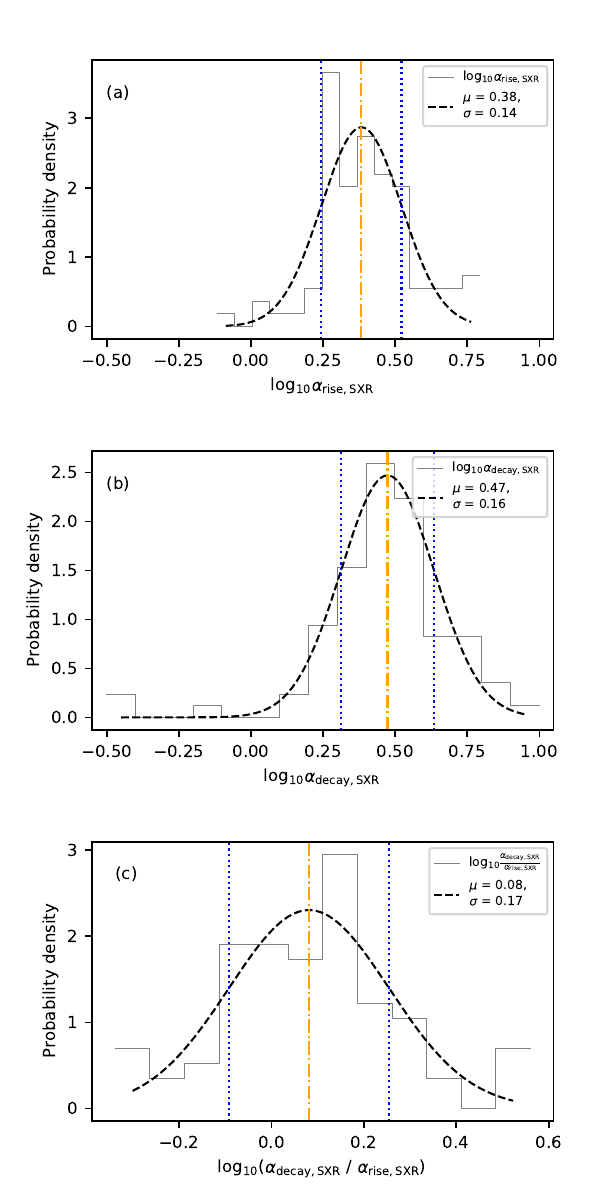}
    \caption{Distributions of spectral indices of SXR spectra at the rise phase (a) and decay phase (b), as well as the index ratio (c). 
    Panel legends and the meaning of line styles are the same as in Fig. \ref{distribution_tp_FMF}.}
    \label{distribution_alpha}
\end{figure}


\noindent We obtain the power-law spectral indices for the SXR spectra based on the sample of 129 flares in categories F, M-F S and M-S. 
The rise and decay phases of the flares are analysed separately. 
Note that we exclude those events with non-positive power-law indices for further statistics. 

\indent The distributions of spectral indices $\alpha_\mathrm{rise,SXR}$ and $\alpha_\mathrm{decay,SXR}$ are shown in 
Fig. \ref{distribution_alpha}a and \ref{distribution_alpha}b. 
We find that the spectral indices for the two phases are distributed in a relatively similar range, mainly from $\sim 1.0 \ (10^{0.0})$ to $\sim 6.3 \ (10^{0.8})$. 
However, their averages present a slight difference, i.e., 
$10^{0.38 \pm 0.14(1\sigma)}$ for the rise phase and $10^{0.47 \pm 0.16(1\sigma)}$ for the decay phase. 

\indent 
After calculating the ratio of the spectral index during the decay phase to that during the rise phase and fitting the probability distribution as a function of the index ratio (in logarithmic scale), as shown in Fig. \ref{distribution_alpha}c, 
we obtain the most probable value of $\log_{10}(\alpha_\mathrm{decay,SXR} / \alpha_\mathrm{rise,SXR})$ being $0.08$, which is a little bit higher than zero. 
Some of the flare events have $\alpha_\mathrm{decay,SXR}$ significantly higher than $\alpha_\mathrm{rise,SXR}$, 
implying that the spectrum for the decay phase of these events tends to be softer than that for the rise phase. 
It implies that for these events, more non-thermal emissions could be generated during the rise phase. 


\subsection{Occurrence frequency distribution of stellar flares} \label{section_occurcence_frequency_distribution_of_stellar_flares}
\begin{figure}
    \centering
    \includegraphics[width=0.5\textwidth]{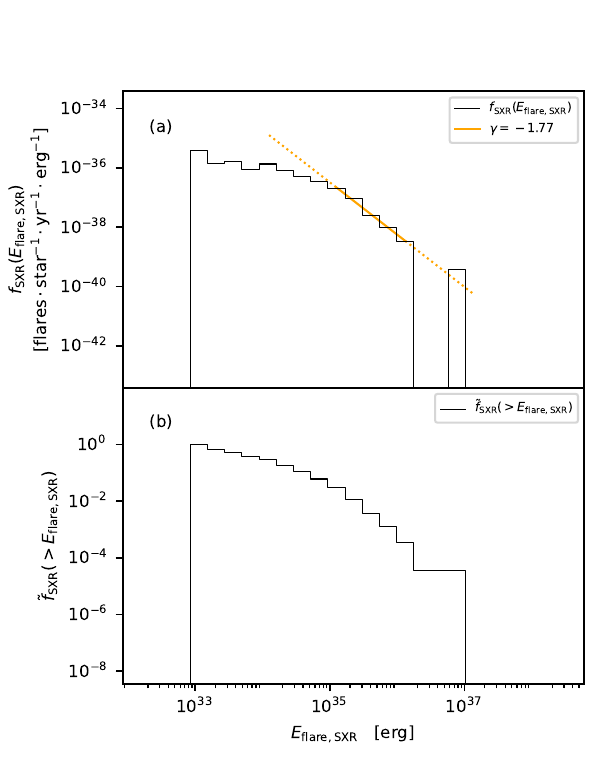}
    \caption{
    (a) Occurrence frequency of flares as a function of flare energy. 
    (b) Normalised cumulative occurrence probability as a function of flare energy. 
    The occurrence frequency as a function of flare energy can be fitted by a power-law function 
    $f_\mathrm{SXR} \propto E_\mathrm{flare,SXR}^\gamma$ in the energy range 
    $(1.68 \times 10^{35},\ 1.76 \times 10^{36}) \ \mathrm{erg}$, as shown with the orange solid line, where the power-law index is $\gamma=-1.77$. 
    The orange dotted lines represent an extension of the fitted line. }
    \label{frequency_distribution}
\end{figure}
\noindent We use the methods mentioned in Section \ref{section_estimation_of_flare_frequencies} to estimate flare occurrence frequencies. 
The results given by equations (\ref{FDF_E}) and (\ref{CFDF_E}) are presented in the two panels of Fig. \ref{frequency_distribution}, respectively. 

\indent The frequency distribution as a function of flare energy, i.e. $f_\mathrm{SXR} (E_\mathrm{flare,SXR})$, is fitted by a power-law function 
for a specific SXR energy range, as shown in Fig. \ref{frequency_distribution}a. 
The fitting gives 
$\mathrm{d} N_\mathrm{flare} / \mathrm{d} E_\mathrm{flare,SXR} \propto E_\mathrm{flare,SXR}^{-1.77}$, 
where we select the energy range from $1.68 \times 10^{35} \ \mathrm{erg}$ to $1.76 \times 10^{36} \ \mathrm{erg}$ beyond which we lack sufficient flares to ensure a reliable fitting \citep{Maehara2012Nat485.478}. 
The fitting result suggests that the average occurrence frequency of stellar flares on each solar-type star is at least 
once in about 2.89 years in the SXR energy range of $10^{34}-10^{35} \ \mathrm{erg}$, 
once in about 16.8 years in the SXR energy range of $10^{35}-10^{36} \ \mathrm{erg}$, 
once in about 98.2 years in the SXR energy range of $10^{36}-10^{37} \ \mathrm{erg}$, and 
once in about 573  years in the SXR energy range of $10^{37}-10^{38} \ \mathrm{erg}$. 

\indent The flare frequency obtained in this work is higher than those at the optical band in previous works \citep[e.g. ][]{Maehara2012Nat485.478,Maehara2015EPS67.59,Shibayama2013ApJS209.5}, and such a difference is mainly due to the following reason. 
As a stellar flare occurs, the radiation at the SXR band often increases significantly in magnitude, while that at the optical band only increases slightly, which thus makes stellar SXR flares to be detected more easily than the optical ones. 
In terms of physical mechanisms of solar flares, generally speaking, it is relatively difficult for the energy released in the corona to be transported to the photosphere \citep[e.g., ][]{Benz2017LRSP14.2,Song2018ApJ867.159}. 

\indent It is interesting that the power-law indices we derived above are quite similar to what have been obtained from different types of stars at different bands \citep[all roughly $-1.8$; e.g. ][]{Shibayama2013ApJS209.5}. 
More interestingly, the spectral indices for stellar flares also resemble with the values for solar activities, i.e., 
$-1.79$ for solar extreme ultraviolet (EUV) nanoflares \citep{Aschwanden2000ApJ535.1047}, 
$-1.74$ for solar SXR microflares \citep{Shimizu1995PASJ47.251}, and 
$-1.53$ for solar hard X-ray (HXR) flares \citep{Crosby1993SoPh143.275}. 
Such a similarity implies that, even though 
solar and stellar flares are from different types of stars, they may share the same physical mechanism, most likely magnetic reconnection \citep[e.g., ][]{Maehara2015EPS67.59,Shibata2016IAUS320.3,Namekata2017ApJ851.91}. 
Thus, the frequency distribution of stellar flares can also serve as a prediction of the probability and frequency of potential superflares occurring on our Sun. 
Meanwhile, the similarity of $\gamma$ obtained at the SXR band and the bolometric value derived at the optical band is a good indication that the energy emitted at the SXR band is nearly a constant fraction of full-band bolometric energy, independent of the magnitude of flares.

\subsection{Completeness} \label{section_completeness}
\noindent The flares with lower flux or shorter duration are usually more difficult to be detected, thus resulting in a selection bias or incomplete sample set. 
In order to check the completeness of our sample, we perform a simulation and generate a series of simulated flare light curves with different $T_\mathrm{duration}$ and $\mathrm{CR}_\mathrm{peak}$ (see Appendix \ref{appendix_completeness_simulation_process}). 
We then analyse these simulated light curves using the same methods as mentions in Sections \ref{section_analysis_of_light_curves} and \ref{section_flare_properties_and_error_analysis}, and obtain their light curve categories and time parameters. 
If a light curve is classified into F, S, R, D or M categories, we regard it as a ``detected event'', otherwise (C, V or B categories) we regard it as an ``undetected event''. 
Then the completeness distribution against $T_\mathrm{duration}$ and $\mathrm{CR}_\mathrm{peak}$ is calculated as 
\begin{align}
    &\text{Completeness} \left( T_\mathrm{duration}, \mathrm{CR}_\mathrm{peak} \right) = \nonumber \\ 
    &\frac{\text{Number of detected events} \left( T_\mathrm{duration}, \mathrm{CR}_\mathrm{peak} \right)}{\text{Number of simulated events} \left( T_\mathrm{duration}, \mathrm{CR}_\mathrm{peak} \right)}, \label{completeness_calculation}
\end{align}
where the number of simulated events at a certain $\left( T_\mathrm{duration},\mathrm{CR}_\mathrm{peak} \right)$ is fixed to be $100$ in this simulation. 

\begin{figure}
    \centering
    \includegraphics[width=0.5\textwidth]{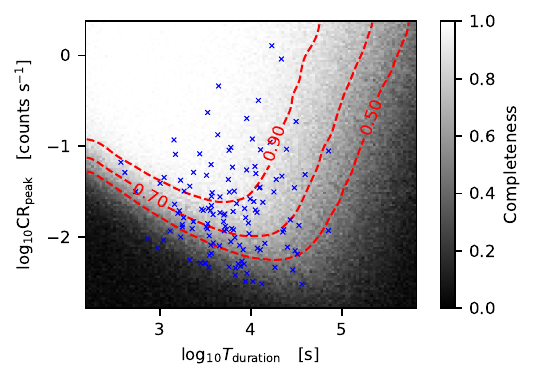}
    \caption{Completeness distribution against $T_\mathrm{duration}$ and $\mathrm{CR}_\mathrm{peak}$. 
    The completeness of the sample set at each $\left(T_\mathrm{duration},\mathrm{CR}_\mathrm{peak}\right)$ is illustrated by the color bar, which is generated from simulated light curves (see Section \ref{section_completeness} and Appendix \ref{appendix_completeness_simulation_process}). 
    Three contours with completeness of $0.50$, $0.70$ and $0.90$ are shown with red dashed lines. 
    The distribution of the real observed flare events in this work is shown with blue crosses. }
    \label{completeness}
\end{figure}

\indent The completeness distribution is shown in Figure \ref{completeness}. 
At each certain $\mathrm{CR}_\mathrm{peak}$, the completeness first rises with $T_\mathrm{duration}$ and then falls. 
The completeness is relatively low when $T_\mathrm{duration}$ is very long because the exposure time could hardly cover the full flare event, thus causing $\mathrm{CR}_\mathrm{q}$ overestimated and then hard to identify a flaring phase (see Section \ref{section_analysis_of_light_curves}). 
For most of the flare events we study, the completeness is higher than $50\%$ even though only a few events have the completeness higher than $90\%$ according to the simulation. 
The histograms of the time and energetic parameters in this work could be affected by the completeness issue, as the low-flux or short-duration flares often have a low completeness, or the probability of detection is low. 
If more low-flux or short-duration flares are detected, the distributions of time and energetic parameters could be more like power-law distributions, like flares on the Sun \citep[e.g., ][]{Kou2020ApJL898.24}. 
Here, we tend to only present the observational data in the histograms because some assumptions that are made in the completeness simulation (see Appendix \ref{appendix_completeness_simulation_process}) have not been verified. 
The simulation results can be used as a guide for determining whether the observational sample is complete but are not intended for further analysis. 

\indent However, it should be addressed that the major conclusions drawn from the flares with the highest peak flux and the longest duration as derived in this work, as well as the correlations among the time, energetic parameters and the occurrence frequency distribution, are not influenced by the completeness issue. 

\section{Summary} \label{section_summary}
\noindent We have made a statistical study for 129 flare events on 103 nearby solar-type stars from Chandra's CSC 2.0 catalog. 
We presented a comprehensive set of methods for identifying and classifying SXR stellar flares, 
and obtained the following main results and conclusions: 
\begin{itemize}
\item[1.] The flare duration at the SXR band ranges from $\sim 10^{2.0}$ to $\sim 10^{4.5} \ \mathrm{s}$, 
with the majority of flares showing a longer duration than typical solar SXR flares. 
The duration of the decay phase occupies most of the entire duration of flares. 

\item[2.] The flare peak luminosity at the SXR band ranges from $\sim 10^{29}$ to $\sim 10^{34} \ \mathrm{erg \ s^{-1}}$ and 
the total flare energy at the SXR band emitted varies from $\sim 10^{33}$ to $\sim 10^{37} \ \mathrm{erg}$, 
which is much larger than the strongest solar flare ever recorded. 
The energy of the strongest stellar flare is found to be $6.0_{-4.7}^{+3.2} \times 10^{37} \ \mathrm{erg}$. 
The majority of flares in our sample are superflares, 3 of which are megaflares (with $E_\mathrm{flare,SXR}$ exceeding $10^{36.2} \ \mathrm{erg}$). 
An upper limit seems to exist in the bolometric energy for a single event, which is $\sim 10^{38}-10^{39} \ \mathrm{erg}$. 

\item[3.] The duration of the flares is essentially independent of the peak luminosity. 
The duration as a function of energy at the SXR band can be best fitted with $T_\mathrm{duration,SXR} \propto E_\mathrm{flare,SXR}^{0.201 \pm 0.024}$, where the power-law index is apparently lower than those derived at optical and NIR bands, 
indicating the distinct radiation processes working at the different wave bands. 

\item[4.] The spectra of some of the flare rise phases are harder than that of the corresponding decay phases, to say, the rise phases of these events could be more non-thermal. 

\item[5.] The frequency distribution of stellar SXR flares as a function of energy can be well fitted by 
$\mathrm{d} N_\mathrm{flare} / \mathrm{d} E_\mathrm{flare,SXR} \propto E_\mathrm{flare,SXR}^{-1.77}$ in the range from 
$1.68 \times 10^{35} \ \mathrm{erg}$ to $1.76 \times 10^{36} \ \mathrm{erg}$. 
The fitted spectral index is similar to the results derived for flares from other types of stars or the Sun. 

\end{itemize}

\noindent Z.H.Z., Z.Q.H., X.C., Z.Y.L. and M.D.D. are funded by National Key R\&D Program of China under grants 2021YFA1600504 and NSFC under grant 12127901. 
Z.H.Z. thanks Dr. Jiacheng Liu, Dr. Jiwei Xie, Lin He, Tong Bao and Guoyin Chen for the very constructive suggestions and helpful comments. 

\bibliography{ref}{}
\bibliographystyle{aasjournal}

\begin{appendix}
\section{A special light curve} \label{appendix_a_special_light_curve}
\begin{figure*}
    \centering
    \includegraphics[width=0.5\textwidth]{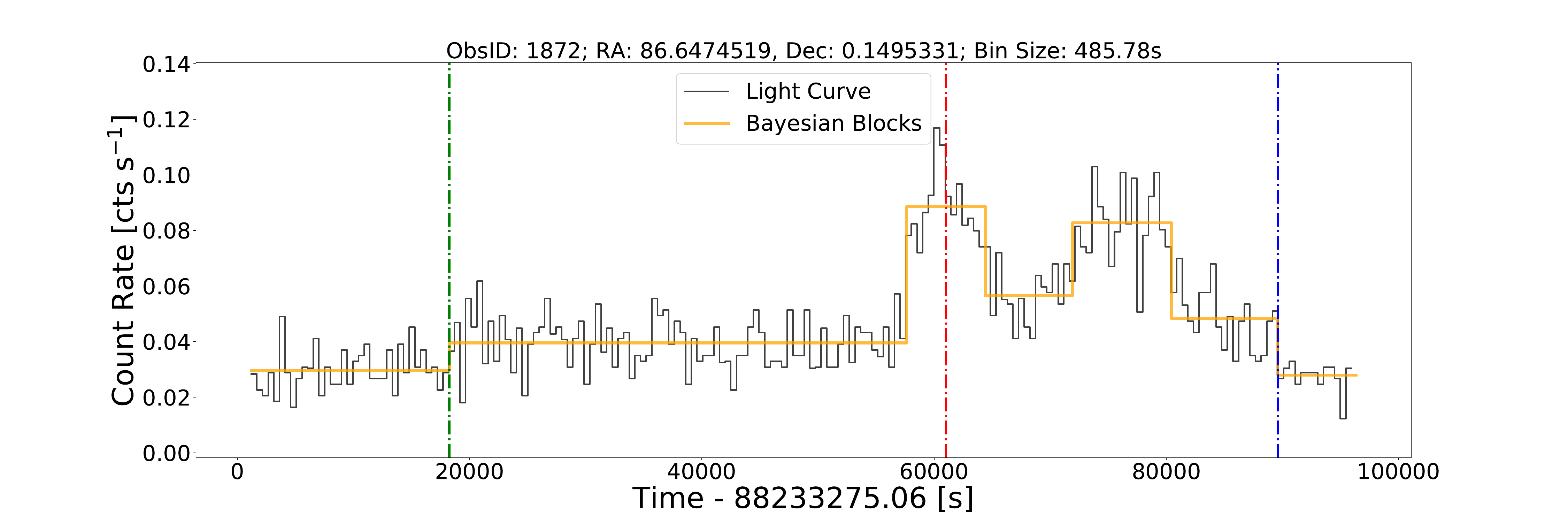}
    \caption{Light curve for the flare event with ObsID 1872, RA 86.64745 and Dec 0.14953, which contains probably two or more unidentified flare events. 
    The meaning of line styles is the same as in Fig. \ref{lc_examples}. } 
    \label{obsid_1872}
\end{figure*}

\section{Completeness simulation process} \label{appendix_completeness_simulation_process}
\begin{figure}
    \centering
    \includegraphics[width=0.5\textwidth]{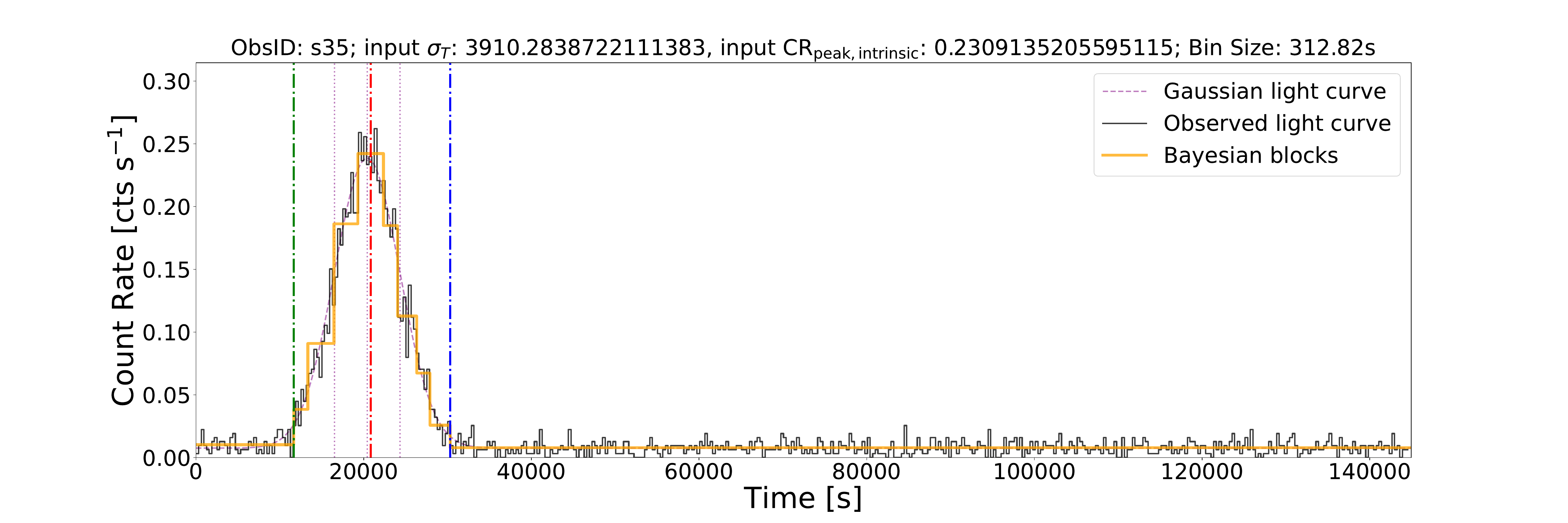}
    \caption{An example of the simulated light curves. The meaning of the black and orange solid lines, as well as the green, red and blue dash-dotted lines is the same as in Fig. \ref{lc_examples}. The Gaussian-shaped light curve is illustrated with a purple dashed line while the center and $1\sigma_T$ deviations of the Gaussian-shaped light curve are illustrated with purple dotted lines. The input $\sigma_T$, $\mathrm{CR}_\mathrm{peak,intrinsic}$ values and the bin size are given above the panel. }
    \label{lc_example_simulated}
\end{figure}

We perform a simulation of light curves in order to check the completeness of our sample. 
The simulation methods are mainly adopted from the Appendix D of \citet{Mossoux2017A&A604.A85}. 
We consider a Gaussian-shaped flare light curve superimposed on a constant level (i.e. $\mathrm{CR}_\mathrm{q,intrinsic}$). 
The input, or ``intrinsic'' light curve (i.e. $\mathrm{CR}_\mathrm{intrinsic}(t)$) can be expressed as 
\begin{equation}
    \mathrm{CR}_\mathrm{intrinsic}(t) = \mathrm{CR}_\mathrm{q,intrinsic} + \mathrm{CR}_\mathrm{peak,intrinsic} \exp \left[ - \frac{1}{2} \left( \frac{t - t_\mathrm{peak,intrinsic}}{\sigma_T} \right)^2 \right], 
\end{equation}
where $\mathrm{CR}_\mathrm{peak,intrinsic}$ is the maximum count rate of the Gaussian component and $\sigma_T$ stands for the standard deviation of the corresponding Gaussian distribution. 
$\sigma_T$ and $\mathrm{CR}_\mathrm{peak,intrinsic}$ are the input values of a simulated light curve. 
Other properties of a simulated light curve include the total ``exposure'' time $T_\mathrm{exposure}$, the quiescent count rate $\mathrm{CR}_\mathrm{q,intrinsic}$ and the peak time $t_\mathrm{peak,intrinsic}$. 
When performing the simulation, we let $\log_{10}T_\mathrm{exposure}$ and $\log_{10}\mathrm{CR}_\mathrm{q,intrinsic}$ obey the normal distribution in logarithmic scale fitted by the observational data, i.e. 
\begin{align}
    \log_{10} T_\mathrm{exposure}    \sim&\ N(\mu =  4.391, \sigma= 0.496),\\
    \log_{10} \mathrm{CR}_\mathrm{q,intrinsic} \sim&\ N(\mu = -2.413, \sigma= 0.587), 
\end{align}
and let $t_\mathrm{peak,intrinisc}$ randomly distribute between $0$ and $T_\mathrm{exposure}$. 
At each $\left(\sigma_T,\mathrm{CR}_\mathrm{peak,intrinsic}\right)$, we simulate $100$ events and calculate the completeness by using equation (\ref{completeness_calculation}). 
Figure \ref{lc_example_simulated} shows an example of the simulated light curves. 

\begin{figure}
    \centering
    \includegraphics[width=0.5\textwidth]{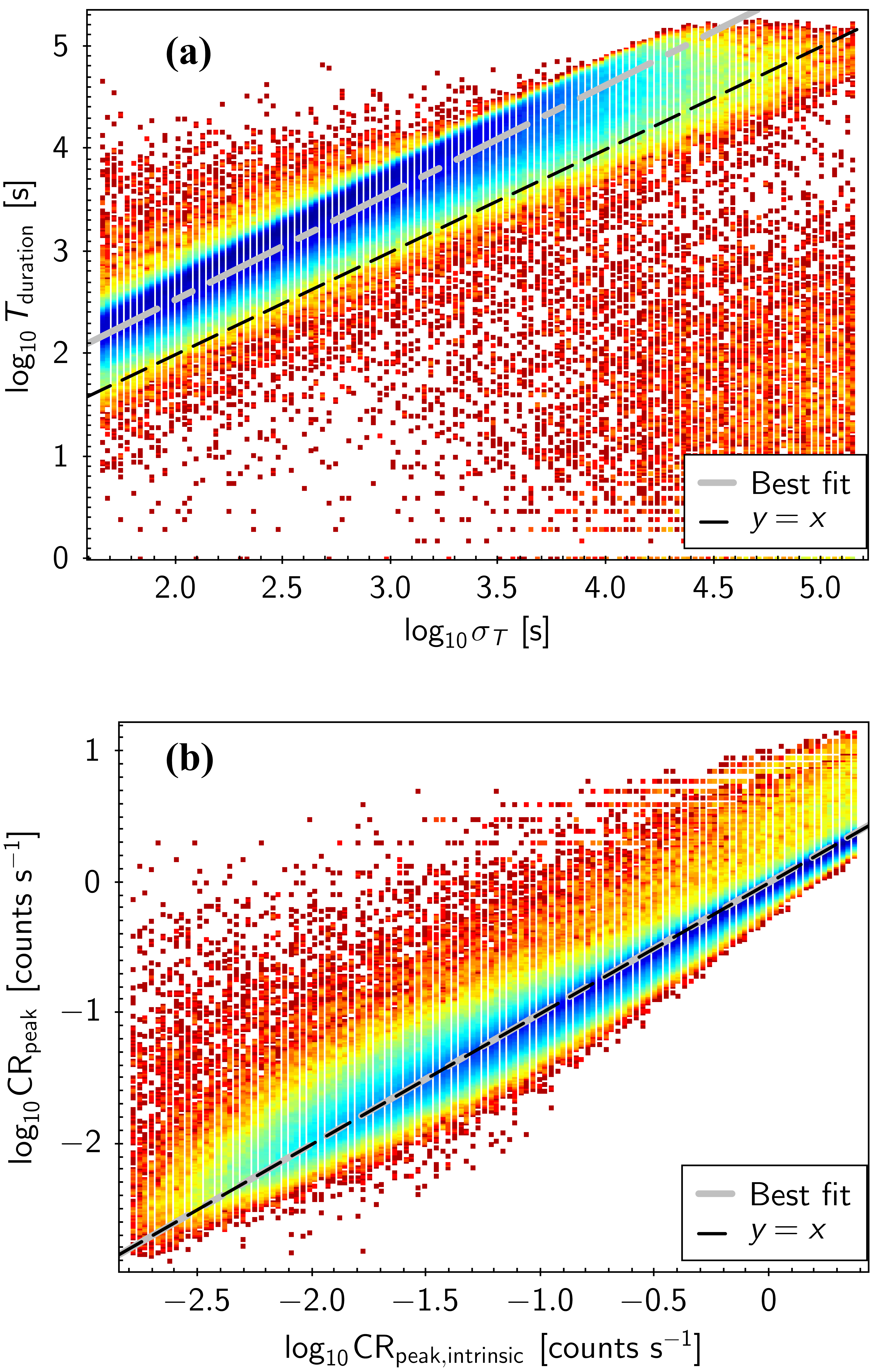}
    \caption{Scatter plots of (a) $T_\mathrm{duration} - \sigma_T$ and (b) $\mathrm{CR}_\mathrm{peak} - \mathrm{CR}_\mathrm{peak,intrinsic}$ of the simulated flare light curves. The color shows the areal density of the sample: the bluer parts are denser than the redder parts. The black dashed lines show reference lines of $y=x$, or where the ``observed'' values equal the ``intrinsic'' values. The best-fit power-law relations are shown in grey dash-dotted lines. When fitting the scatter plots, abnormally diffuse data points are excluded. }
    \label{correlation_output_input}
\end{figure}

\indent It's necessary to obtain the correlations between $\left( \sigma_T, \mathrm{CR}_\mathrm{peak,intrinsic} \right)$ and $\left( T_\mathrm{duration}, \mathrm{CR}_\mathrm{peak} \right)$, or between the intrinsic and observed properties of the light curves when we check the completeness of our observational data. 
We plot the scatters of $T_\mathrm{duration} - \sigma_T$ and $\mathrm{CR}_\mathrm{peak} - \mathrm{CR}_\mathrm{peak,intrinsic}$ of the simulated data, as shown in Figure \ref{correlation_output_input}, and fit the scatters with power-law relations. 

The best fits give
\begin{align}
    T_\mathrm{duration}       =&\ 2.7956 \left( \sigma_T \right)^{1.0443}, \\
    \mathrm{CR}_\mathrm{peak} =&\ 1.0076 \left( \mathrm{CR}_\mathrm{peak,intrinsic} \right)^{1.0012}, 
\end{align}
which are employed to obtain the completeness distribution of observational data when plotting Figure \ref{completeness}. 
The best fits suggest that $\mathrm{CR}_\mathrm{peak}$ is nearly the same as $\mathrm{CR}_\mathrm{peak,intrinsic}$, while $T_\mathrm{duration}$ is about $2.8$ times of $\sigma_T$. 
In this case, the full width at half maximum (FWHM) of the Gaussian-shaped light curve 
\begin{equation}
    T_\mathrm{FWHM} = 2 \sqrt{2 \ln{2}} \sigma_T \approx 2.355 \sigma_T 
\end{equation}
is relatively close to the duration given by the Bayesian blocks, i.e. $T_\mathrm{duration}$ (see Sections \ref{section_analysis_of_light_curves} and \ref{section_flare_properties_and_error_analysis}).

\end{appendix}

\end{document}